\documentclass[useAMS,twocolumn]{mn2e}
\usepackage{amssymb}
\usepackage{graphicx}

\newcommand       \be           {\begin{equation}}
\newcommand       \ee           {\end{equation}}
\newcommand       \ba           {\begin{eqnarray}}
\newcommand       \ea           {\end{eqnarray}}
\newcommand       \grad         {\nabla}

\def\lesssim{\mathrel{\hbox{\rlap{\hbox{\lower4pt\hbox{$\sim$}}}\hbox{$<$}}}}
\def\gtrsim{\mathrel{\hbox{\rlap{\hbox{\lower4pt\hbox{$\sim$}}}\hbox{$>$}}}}

\title[Hot accretion flow with anisotropic thermal conduction]
{Effects of anisotropic thermal conduction on wind properties in hot
accretion flow}
\author[Bu, et al.]
{De-Fu Bu$^{1}$\thanks{dfbu@shao.ac.cn}; Mao-Chun
Wu$^2$\thanks{maochun@ustc.edu.cn};  Ye-Fei Yuan$^{2}$ \\
$^{1}$Key Laboratory for Research in Galaxies and Cosmology,
Shanghai Astronomical Observatory, \\ Chinese Academy of Sciences,
80 Nandan Road, Shanghai, 200030, China\\ $^{2}$Key Laboratory for
Research in Galaxies and Cosmology CAS, Department of Astronomy,
\\University of Science and Technology of China, Hefei, Anhui
230026, China\\}
\begin{document}

\maketitle

\label{firstpage}

\begin{abstract}
Previous works have clearly shown the existence of winds from black
hole hot accretion flow and investigated their detailed properties.
In extremely low accretion rate systems, the collisional mean-free
path of electrons is large compared with the length-scale of the
system, thus thermal conduction is dynamically important. When the
magnetic field is present, the thermal conduction is anisotropic and
energy transport is along magnetic field lines. In this paper, we
study the effects of anisotropic thermal conduction on the wind
production in hot accretion flows by performing two-dimensional
magnetohydrodynamic simulations. We find that thermal conduction has
only moderate effects on the mass flux of wind. But the energy flux
of wind can be increased by a factor of $\sim 10$ due to the
increase of wind velocity when thermal conduction is included. The
increase of wind velocity is because of the increase of driving
forces (e.g. gas pressure gradient force and centrifugal force) when
thermal conduction is included. This result demonstrates that
thermal conduction plays an important role in determining the
properties of wind.

\end{abstract}


\section{Introduction}
Hot accretion flow such as advection-dominated accretion flow (ADAF;
Narayan \& Yi 1994, 1995; Abramowicz et al. 1995) is interesting
because it can be used to model the low-luminosity active galactic
nuclei (LLAGNs), which are the majority of galaxies at least in the
nearby universe, and the hard/quiescent states of black hole X-ray
binaries (see Yuan \& Narayan 2014 for the latest review of current
theoretical understanding of hot accretion flow and its various
astrophysical applications).

Numerical simulations of hot accretion flow show that the mass
inflow rate (see Equation (\ref{inflowrate}) for definition)
decreases with decreasing radius $\dot {M}_{\rm in} (r) \propto r^s$
with $s\sim 0.5-1$ (e.g. Stone et al. 1999; Igumenshchev \&
Abramowicz 1999, 2000; Hawley \& Balbus 2002; Pang et al. 2011; Yuan
et al. 2012a; Bu et al. 2013). Especially Yuan et al. (2012b) show
that the inward decrease of the accretion rate is due to the
significant mass loss via wind (see also Narayan et al. 2012; Li et
al. 2013). This conclusion is soon confirmed by the 3 million
seconds {\it Chandra} observations of the accretion flow around the
super-massive black hole in the Galactic Center, combined with the
modeling to the detected iron emission lines (Wang et al. 2013).
Begelman (2012) and Gu (2015) address the question of why wind
exists. The detailed properties of wind such as the mass flux,
angular distribution, terminal velocity, and fluxes of energy and
momentum, have been studied in Yuan et al. (2015) (see also Sadowski
et al. 2016) by following the trajectories of the fluid particles.
Bu et al. (2016) show that wind production can only occur within the
Bondi radius of the accretion flow.

All the simulations mentioned above have neglected the effects of
thermal conduction. However, thermal conduction is very important
when the accretion rate is very low.  In low accretion rate systems,
the electron mean free path is very large. In this case, thermal
conduction can have a significant influence on the dynamics of the
accretion flow (Quataert 2004; Johnson \& Quataert 2007), resulting
in the transport of thermal energy from the inner (hotter) to the
outer (cooler) regions. If the energy flux carried by thermal
conduction is substantial, the temperature of the gas in the outer
regions can be increased above the virial temperature. Thus, gas in
the outer regions is able to escape from the gravitational potential
of the central black hole and form outflows, significantly
decreasing the mass accretion rate (Tanaka \& Menou 2006; Johnson \&
Quataert 2007; Sharma et al. 2008; Bu et al. 2011).

If the mean free path of electron is much larger than its
gyro-radius, conduction will be anisotropic and along magnetic field
lines (Balbus 2000, 2001; Parrish \& Stone 2005, 2007; Quataert
2008; Foucart 2016). In many LLAGNs, the electron mean free path is
much larger than electron gyro-radius (see Tanaka \& Menou 2006).
Let's take the accretion flow in Galactic Center as an example to
compare the electron mean free path to their gyro-radius. From
\emph{Chandra} observations of the accretion flow at the Galactic
Center, Sgr A*, the electron mean-free path is estimated to be
$0.02-1.3$ times the Bondi radius, i.e.,  the electron mean free
path $l \sim 1.3 \times 10^{17} \rm cm$ (Tanaka \& Menou 2006). The
gyro-radius of electrons is $R_{\rm gyro}=m_e v_{\rm {th}} c/qB \sim
4\times 10^5\sqrt{\beta/n}$, where $m_e$ is electron mass, $v_{\rm
th}$ is thermal speed of electron, $c$ is speed of light, $q$ is
electron charge, $B$ is magnetic field, $\beta$ is the ratio between
gas pressure and magnetic pressure, $n$ is number density of
electrons. At $10^5 R_{\rm s}$, $n=100$, $R_{\rm gyro} \sim 10^5 \rm
cm $ if we assume $\beta \sim 10$; Therefore, it is clear that
electron mean free path is much larger than its gyro-radius. For the
accretion flow at Galactic Center $n \sim r^{-1/2}$, so $R_{\rm
gyro} \propto r^{1/4}$ if we assume $\beta$ is almost constant in
the accretion flow. But the electron mean free path $l \propto
r^{-3/2}$ (Tanaka \& Menou 2006). As a result, the electron mean
free path will become even larger than gyro-radius when approaching
the central black hole event horizon.

In this paper, we study the effects of anisotropic thermal
conduction on the properties of hot accretion flow by performing
two-dimensional magnetohydrodynamic (MHD) simulations. We especially
pay attention to the wind properties, such as the mass, momentum and
energy fluxes of wind launched from the accretion flow. These
quantities are essential for the study of AGN feedback since winds
play an important role in the feedback process. For example, the two
gamma-ray Bubbles observed by the \emph{Fermi}-LAT below and above
the Galactic plane (Su et al. 2010) may be inflated by the wind from
the hot accretion flow in our Galactic Center (Mou et al. 2014). In
\S 2, we will describe the basic equations and the simulation
method. In \S 3, we will present the results. We discuss and
summarize our results in \S4.

\section{Numerical method and models}
\subsection{Numerical method}
We adopt two-dimensional, spherical coordinates ($r,\theta,\phi$)
and assume axisymmetric ($\partial/\partial \phi=0$) . We use the
Zeus-2D code (Stone \& Norman 1992a, 1992b) to solve the MHD
equations with anisotropic thermal conduction:

\begin{equation}
 \frac{d\rho}{dt} + \rho \nabla \cdot {\bf v} = 0,
\end{equation}
\begin{equation}
 \rho \frac{d{\bf v}}{dt} = -\nabla p - \rho \nabla \Phi +
 \frac{1}{4\pi}(\nabla \times {\bf B}) \times {\bf B}
\end{equation}
\begin{equation}
 \rho \frac{d(e/\rho)}{dt} = -p\nabla \cdot {\bf v} - \nabla \cdot {\bf
 Q} + \eta {\bf
 J}^{2}
\end{equation}
\begin{equation}
 \frac{\partial {\bf B}}{\partial t} = \nabla \times ({\bf v} \times {\bf B}
- \eta {\bf J})
\end{equation}

\begin{equation}
  {\bf Q } =  -\chi {\bf \hat{b} (\hat{b} \cdot \nabla)} {\bf T}
\end{equation}
Here, $\rho$ is the mass density, $\bf v$ is the velocity, $p$ is
the gas pressure, ${\bf J}=c({\bf \grad \times B})/4\pi$ is the
current density, $\Phi$ is the gravitational potential, $\bf B$ is
the magnetic field, $e=p/(\gamma-1)$ is the internal energy (where
$\gamma$ is the adiabatic index, we set $\gamma=5/3$), ${\bf
\hat{b}} = \bf B/|\bf B|$ is the unit vector in the direction of the
magnetic field, $\eta$ is the explicit resistivity, $T$ is the gas
temperature, ${\bf Q}$ is the heat flux along field lines and $\chi$
is the thermal diffusivity.

The final terms in Equations (3) and (4) are the magnetic heating
and dissipation rate mediated by a finite resistivity $\eta$. Since
the energy equation here is actually an internal energy equation,
numerical reconnection inevitably results in loss of energy from the
system. By adding the anomalous resistivity, the energy loss can be
captured in the form of heating in the current sheet (Stone \&
Pringle 2001). The exact form of $\eta$ is same as that used by
Stone \& Pringle (2001).

We use the pseudo-Newtonian potential to mimic the general
relativistic effects, $\Phi=-GM_{\rm BH}/(r-R \rm s)$, where $G$ is
the gravitational constant and $M_{\rm BH}$ is the mass of the
central black hole. The self gravity of the gas is neglected. In
this paper, we set $GM_{\rm BH}=R\rm s=1$. We use $R\rm s$ to
normalize the length-scale. Time is in unit of the Keplerian orbital
time at $100R \rm s$. The black hole in our Galactic Center has a
mass $M_{\rm BH} \approx 4 \times 10^6 M_\odot$, $M_\odot$ is the
solar mass. For the black hole in our Galactic Center, the length
unit ($R\rm s$) is $\sim 10^{12}$ cm. The orbital time at $100R \rm
s$ is $\sim 3.8 \times 10^5$ seconds.

Following Sharma et al. (2008), we assume $\kappa = \chi T/p $ which
has the dimensions of a diffusion coefficient ($\rm cm^2 s^{-1}$).
In non-relativistic theory, $\kappa \propto c_s^2 \tau_{\rm R}$,
with $c_s$ is the sound speed and $\tau_{\rm R}$ is the effective
mean-free-time due to wave-particle scatterings (Foucart et al.
2016). A nature scale for $\tau_{\rm R}$ is the dynamical time
$\sqrt{r^3/(G M_{\rm BH})}$. For the hot accretion flow such as Sgr
$\rm A^*$, the temperature of gas is almost virial; therefore, the
sound speed is proportional to the Keplerian rotational velocity
$c_s \propto v_k \propto r^{-1/2}$. Thus, for hot accretion flow,
$\kappa \propto r^{1/2}$. In this paper, we assume that $ \kappa
\equiv \alpha_c (G M_{\rm BH} r)^{1/2}$. The precise value of
$\alpha_c$, which depends on the microphysical processes, is
difficult to calculate. Motivated by the results of Sharma et al.
(2006) and Sharma et al. (2007) on the microphysics in collisionless
accretion flows, we take $\alpha_c=0.2$. This value is much smaller
than the maximum free-streaming thermal conductivity that could be
obtained if the electrons are virial.

\subsection{Initial conditions and numerical settings}
Following Stone et al. (1999), the initial state is an equilibrium
torus with constant specific angular momentum which is given by
(Papaloizou \& Pringle 1984),
\begin{equation}
 \frac{p}{\rho} = \frac{GM_{\rm BH}}{(n+1)R_{0}} \left[ \frac{R_{0}}{r} - \frac{1}{2}
\left( \frac{R_{0}}{r \sin \theta} \right)^{2} - \frac{1}{2d}
\right] .
\end{equation}
Here, $R_0$ is the radius of the torus center (density maximum), $n
= (\gamma -1 )^{-1}$ is the polytropic index and $d=1.5$ is the
distortion of the equilibrium torus. We assume that at the torus
center $\rho=1$ and the torus is embedded in a low-density medium
$\rho_m=10^{-4}$.

The computational domain is from $R_{ \rm in} =1.3$ or $2 R_{\rm s}$
to $R_{\rm out}$ = 400 $R_{\rm s}$ in radial direction and from
$\theta = 0^\circ $ to $\theta = 180^\circ $ in angular direction.
The radial grids are logarithmically spaced (grid spacing ${\rm d}r
\propto r$). In angular direction, the grids are uniformly spaced.
The axisymmetric boundary conditions are used at $\theta=0^\circ$
and $180^\circ$. We adopt outflow boundary conditions at both the
inner and outer radial boundaries. The standard resolution is $147
\times 88$. Several modifications to the code were required for the
simulations, including implementing the pseudo-Newtonian potential,
adding the anomalous resistivity and thermal conduction. The thermal
conduction is implemented by using the method based on limiters
(Sharma \& Hammett 2007), this method can guarantee the heat always
flows from hot regions to cold regions.

\begin{figure}

\begin{center}
\includegraphics[scale=0.28]{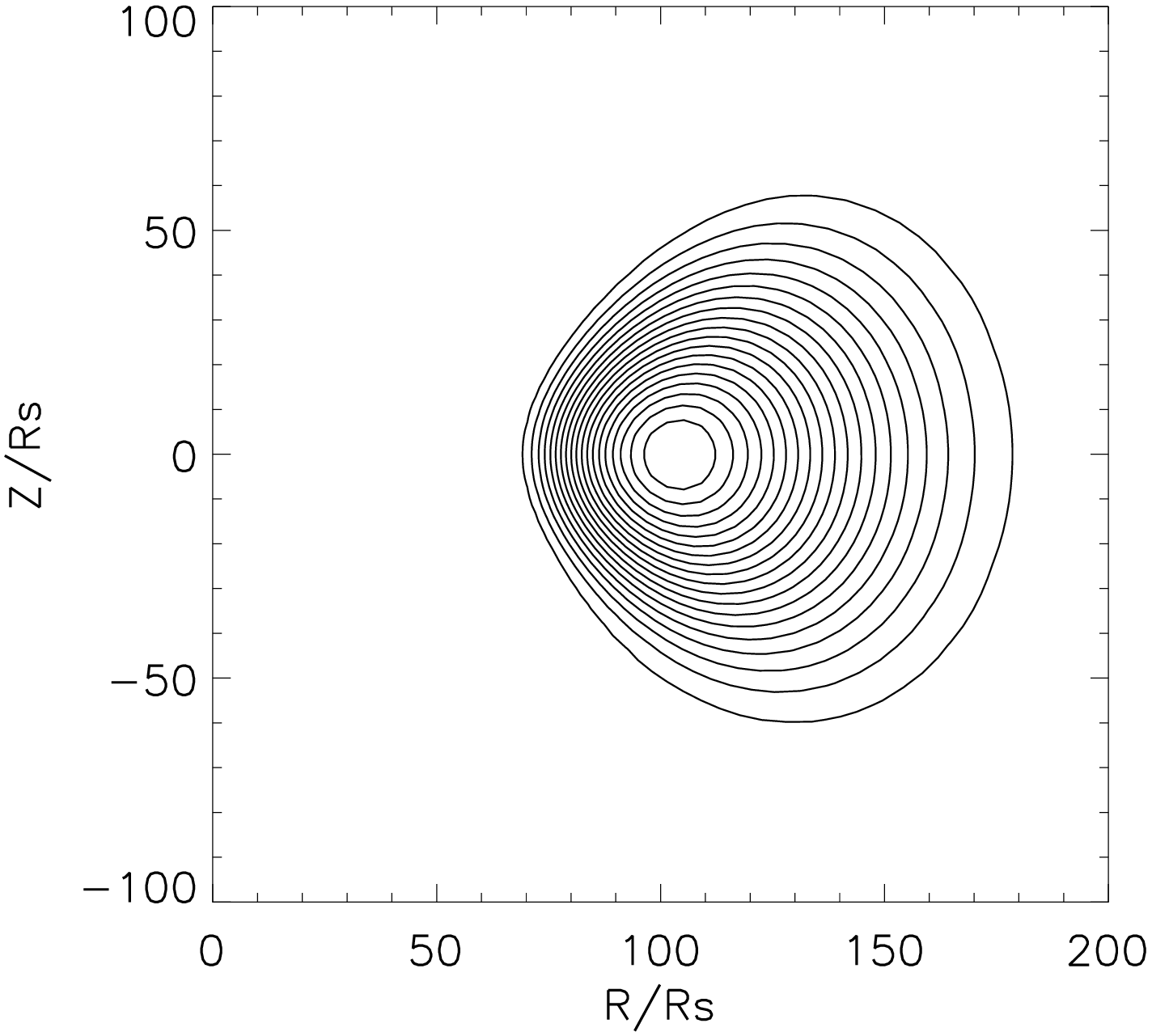}\hspace*{0.5cm}
\includegraphics[scale=0.28]{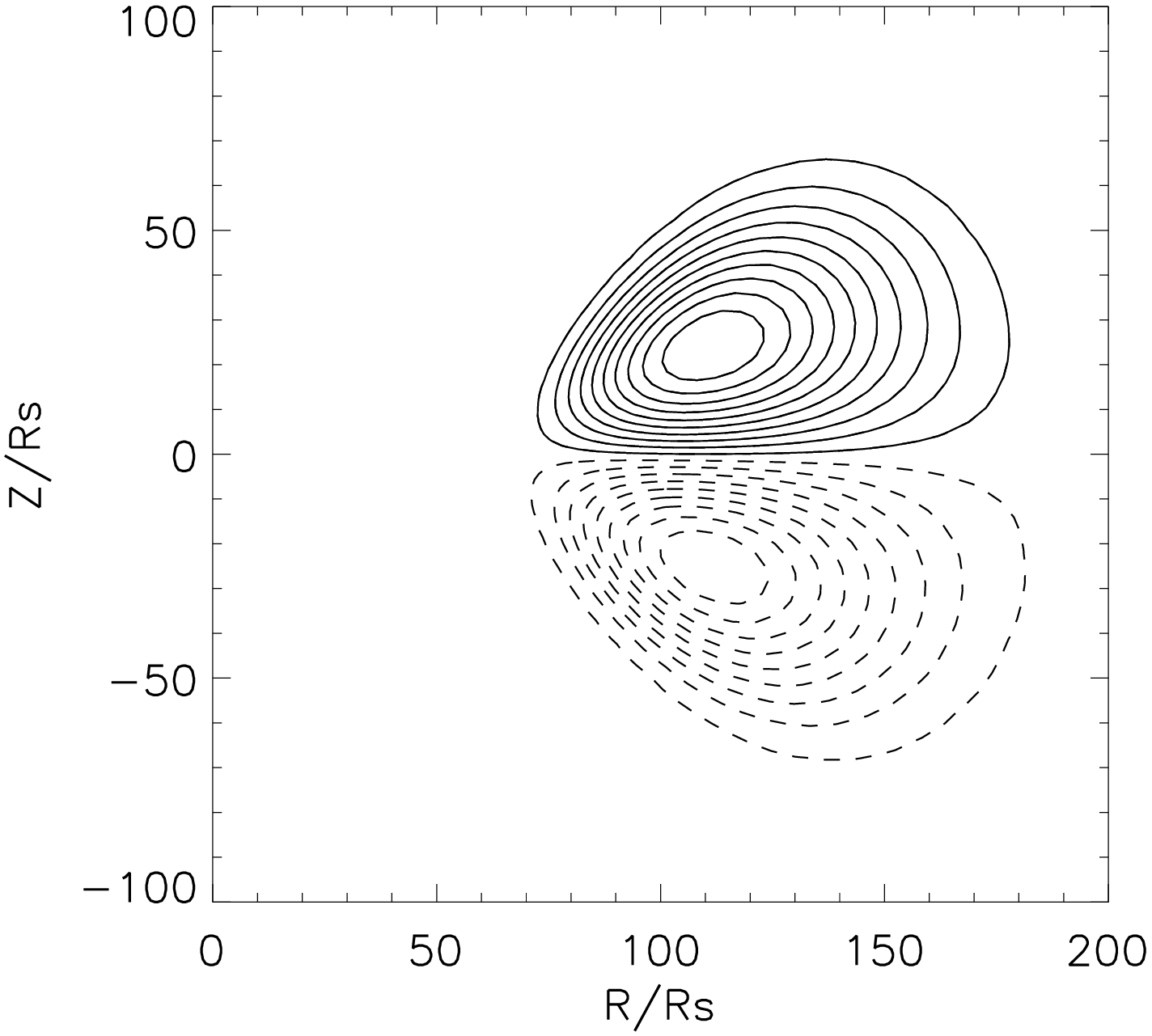}
\hspace*{8cm}

\caption{Initial magnetic field configurations for model A series
(left panel) and B series (right panel). Solid and dashed lines
indicate field polarity: solid lines denote current into the page,
dashed lines denote current out of the page.
 } \label{Fig:field}
\end{center}
\end{figure}

\subsection {Models}
Due to the fact that thermal conduction is transported along
magnetic field lines, if the magnetic field is much ordered, we
expect energy can be transferred from inner to outer region which
will affect the dynamics of the flow significantly. If the magnetic
field is much tangled, we expect the energy can not be transferred
to large distance which may result in small effects on the dynamics
of the flow. Motivated by this point, we explore two different
magnetic configuration: a large scale ordered magnetic field and a
relative small scale tangled magnetic field to examine the effects
of thermal conduction on hot accretion flow.

The magnetic field which threads the torus initially is generated by
a vector potential, i.e. $\bf{B}=\nabla \times \bf A $. Initializing
the field in this way guarantees that it will be divergence-free. We
take $\bf A$ to be purely azimuthal. In model A series, we assume
$\bf A_\phi=\rho^2/\beta_0$ and $\beta_0=100$. This will generate a
dipolar field (see the left panel of Figure \ref{Fig:field}). With
the value of $\beta_0=100$, when the flow achieves steady state, the
field is much ordered. In model B series, we assume the filed is
quadruolar with $\bf A_\phi=\rho^2/\beta_0 r \cos \theta$ (see the
right panel of Figure \ref{Fig:field}). For the quadrupolar filed,
the magnetic loops below and above the equatorial plane have
opposite polarity. Thus, magnetic reconnection is very strong and
finally the magnetic field is prone to be tangled.

\begin{table} \caption{Simulation parameters }
\begin{tabular}{ccccc}
\hline \hline
Name & Field topology & $\alpha_c$ & $\beta_0$ & $t_{f}^\star$\\
\hline
A0  &  dipole      & $0$ & $100$ & $4$ \\
A1 &  dipole      & $0.2$ & $100$ & $4$ \\
B0  &  quadrupole  & $0$ & $50$ & $10$ \\
B1 &  quadrupole  & $0.2$ & $50$ & $10$ \\

\hline
\end{tabular}

$^\star$ Final time at which each simulation is stopped (all times
in this paper are reported in units of the orbital time at $R_0=100
R_{\rm s}$.)

\end{table}

Table 1 lists the main parameters in all models presented here,
initial magnetic field topology, plasmas parameter $\beta_0$ ,
thermal conductivity $\alpha_c$ and final time $t_f$ at which each
simulation is stopped (all times in this paper are reported in units
of the orbital time at $R_0=100 R_s$.)

\section{Results}
We analyze the properties of hot accretion flow at the quasi-steady
state, i.e., the net accretion rate is independent of radius. The
angle integrated mass accretion inflow and outflow rates, $\dot
{M}_{\rm in}$ and $\dot {M}_{\rm out}$, are defined as follows,

\begin{equation}
 \dot{M}_{\rm in}(r) = 2\pi r^{2} \int_{0}^{\pi} \rho \min(v_{r},0)
   \sin \theta \rm d\theta,
   \label{inflowrate}
\end{equation}
\begin{equation}
 \dot{M}_{\rm out}(r) = 2\pi r^{2} \int_{0}^{\pi} \rho \max(v_{r},0)
    \sin \theta \rm d\theta,
    \label{outflowrate}
\end{equation}
and the net mass accretion rate is,
\begin{equation}
\dot{M}_{\rm acc}(r)=\dot{M}_{\rm in}(r)+\dot{M}_{\rm out}(r).
\label{netrate}\end{equation} Note that the above rates are obtained
by time-averaging the integrals rather than integrating the time
averages.

In numerical simulations,  the net mass accretion rate (see Equation
(\ref{netrate})) is always used as a diagnostic to see whether a
quasi-steady state is achieved (e.g. Stone et al. 1999). If a
quasi-steady state is achieved, the net mass accretion rate is
almost a constant of radius. From Figure \ref{Fig:accretionrateA}
(dotted lines), we can see that the net mass accretion rate is
almost a constant of radius. Therefore, the simulations in this
paper have reached a quasi-steady state. In the MHD simulations,
there is turbulence induced by the magneto-rotational instability.
Therefore, the physical quantities always vary with time around
their mean values. As an example, Figure 3 shows the mass flux of
outflow at $r=20R\rm s$ in model A1. It is clear that after t=1.2
orbits, a quasi-steady state is achieved, the mass flux of outflow
oscillates with very small amplitude around its mean value.

\subsection{Dipolar field models}

\begin{figure}
\begin{center}
\includegraphics[scale=0.45]{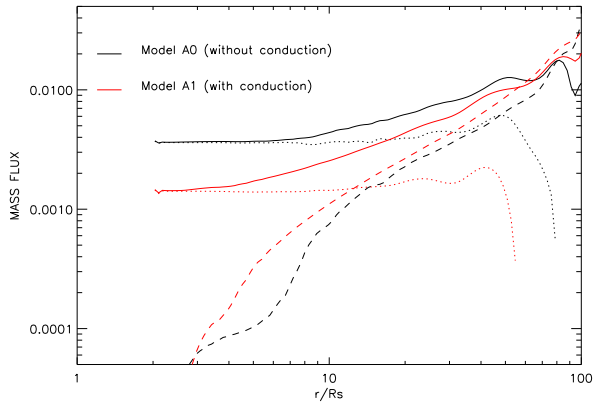}

\caption{ The radial profiles of the time-averaged (from $t=1.5$ to
1.6 orbits) and angle integrated mass inflow rate $\dot{M}_{\rm in}$
(solid line), outflow rate $\dot{M}_{\rm out}$ (dashed line), and
the net rate $\dot{M}_{\rm acc}$ (dotted line) in model A0 (black
lines) and A1 (red lines). \label{Fig:accretionrateA}}
\end{center}
\end{figure}

\begin{figure}
\begin{center}
\includegraphics[scale=0.45]{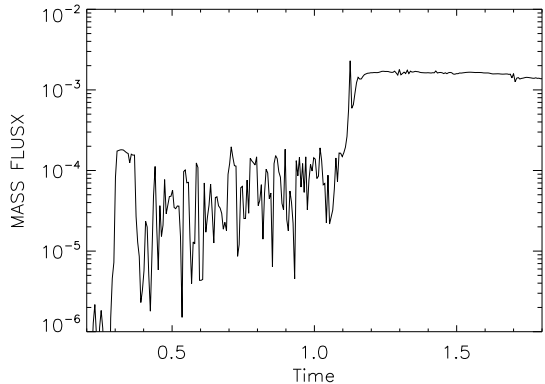}

\caption{ Time evolution of the mass outflow rate at $r=20R \rm s$
calculated by Equation (\ref{outflowrate}) in model A1.
\label{Fig:timerate}}
\end{center}
\end{figure}

At first we analyze model A series, all the data are time averaged
from $t=1.5$ to 1.6 orbits except the snapshot data is at time $t=
1.6$ orbits. Figure \ref{Fig:accretionrateA} shows the radial
profiles of the time-averaged mass inflow rate $\dot{M}_{\rm in}$,
outflow rate $\dot{M}_{\rm out}$ and net rate $\dot{M}_{\rm acc}$ in
model A0 (without thermal conduction, $\alpha_c $= 0.0) and A1 (with
thermal conduction, $\alpha_c $= 0.2). From this figure, it is clear
that the mass inflow rate in both models decreases inwards,
consistent with those found in previous works (see review in Yuan et
al. 2012a). The mass inflow rate in model A1 decreases much quicker
towards the black hole than model A0 and the net mass accretion rate
in model A1  is about half of that in model A0. This is because the
mass outflow rate in model A1 is larger than that in model A0. Note
that we count all the gas with positive velocity as the mass outflow
rate (see Equation (\ref{outflowrate})), i.e., the mass outflow rate
includes both real outflow (wind) and gas which is doing turbulent
motions. Thus the crucial issue to quantitatively study the strength
of wind is  to get rid of the contamination of turbulence. This is
achieved in Yuan et al. (2015) by using a trajectory approach. Based
on the general relativistic magnetohydrodynamic (GRMHD) simulation
data of hot accretion flow (without thermal conduction), they
convincingly show that the mass flux of real outflow is $\sim 60\%$
of the outflow rate calculated by Equation (\ref{outflowrate}).

\begin{figure}
\begin{center}
\includegraphics[scale=0.28]{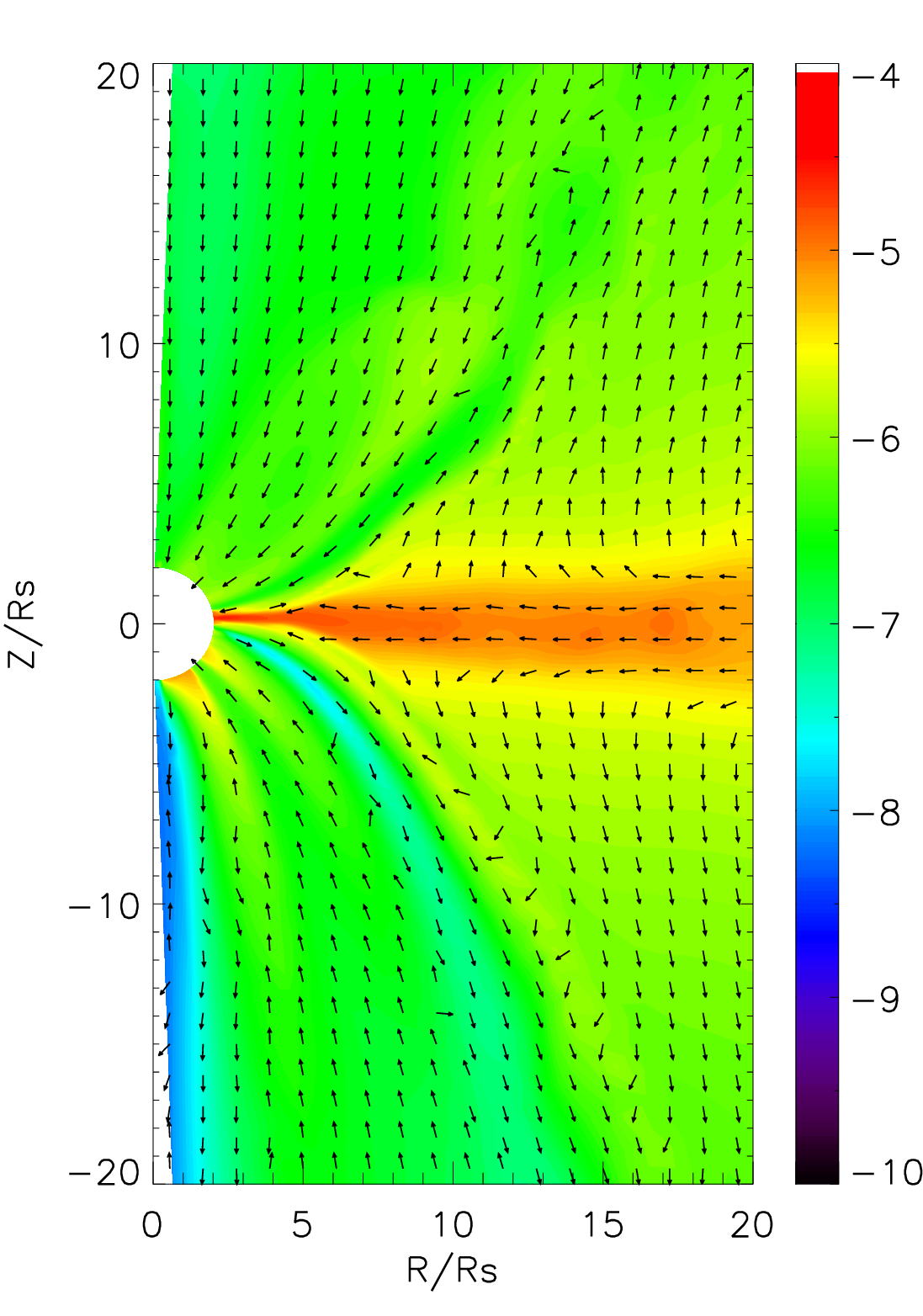}\hspace*{0.5cm}
\includegraphics[scale=0.28]{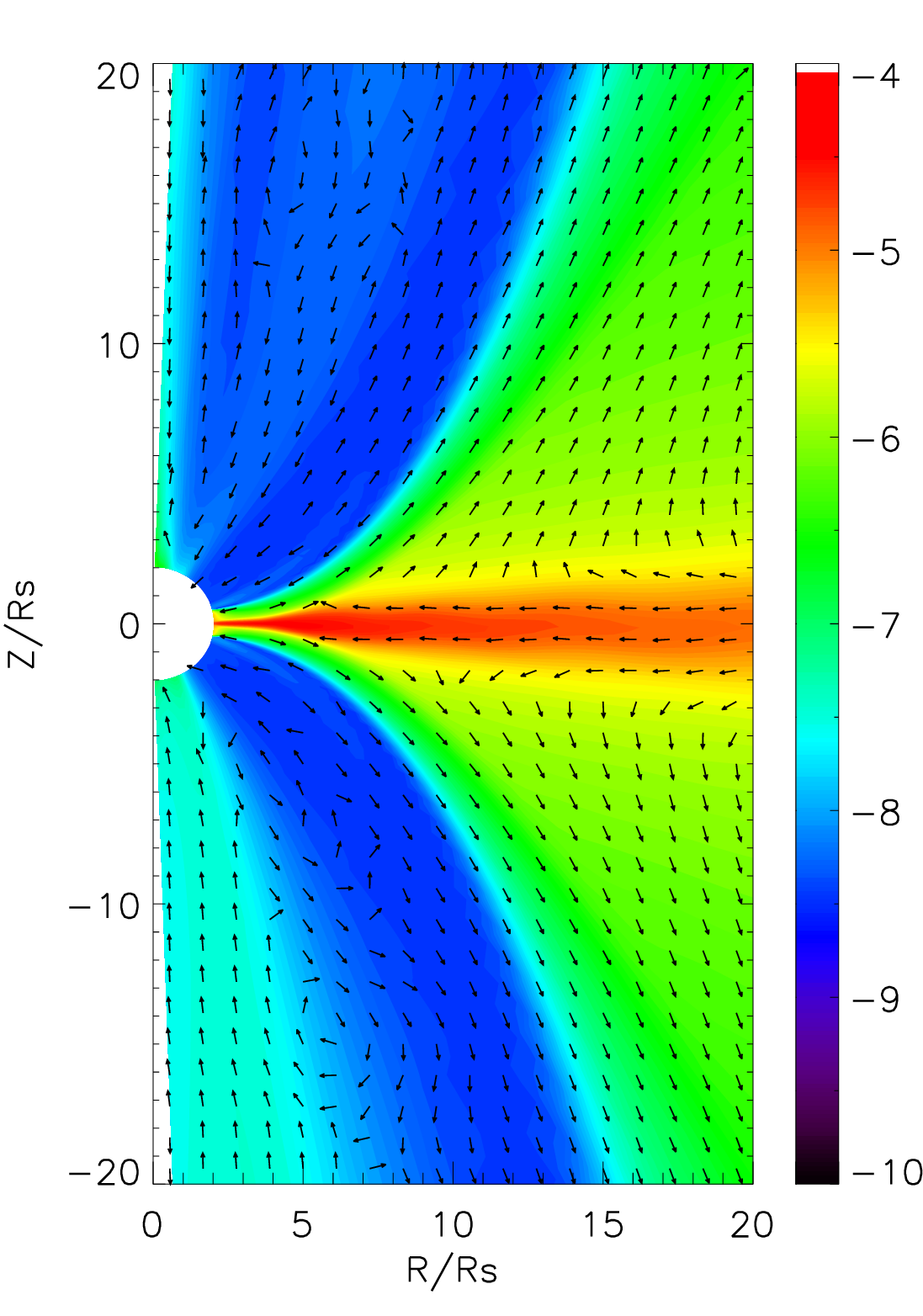}

\caption{Time-averaged (from $t=1.5$ to 1.6 orbits) density and
velocity. Colors show the logarithm density. Arrows show the
direction of velocity ($\bf v/|\bf v|$).  The left pane is for model
A0 (without conduction). The right panel is for model A1 (with
conduction). \label{Fig:velocityfieldA}}
\end{center}
\end{figure}

While it is essential to use the trajectory approach to calculate
the properties of wind such as the mass flux, as shown by Yuan et
al. (2015), this approach is technically complicated and the
calculation is very time-consuming. Since the aim of the present
work is to investigate the effect of thermal conduction, to
qualitatively understand the above-mentioned different results
between models A0 and A1, we find it is enough to simply use the
time-averaged method often adopted in literature. The time-averaged
density contour and velocity field are shown in Figure
\ref{Fig:velocityfieldA}. This figure shows the distribution of
density contour over-plotted by the poloidal velocity field within R
= 20 $R_{\rm s}$ of models A0 (left panel) and A1 (right panel). In
both models, the flow has two components: low temperature high
density main disc body around the equator and high temperature low
density corona sandwiching it. The main disc body is the inflow
region (Sadowski et al. 2013; Yuan et al. 2015). We find that the
wind region in model A1 is larger than that in model A0. For
example, in model A0, it is inflow in the regions $26^\circ <\theta
< 45^\circ $ and $135^\circ <\theta < 154^\circ$. But in model A1,
gas in these two regions becomes wind. The broader wind region in
model A1 results in stronger wind and more rapid inward decrease of
the inflow rate.

\begin{figure}
\begin{center}
\includegraphics[scale=0.28]{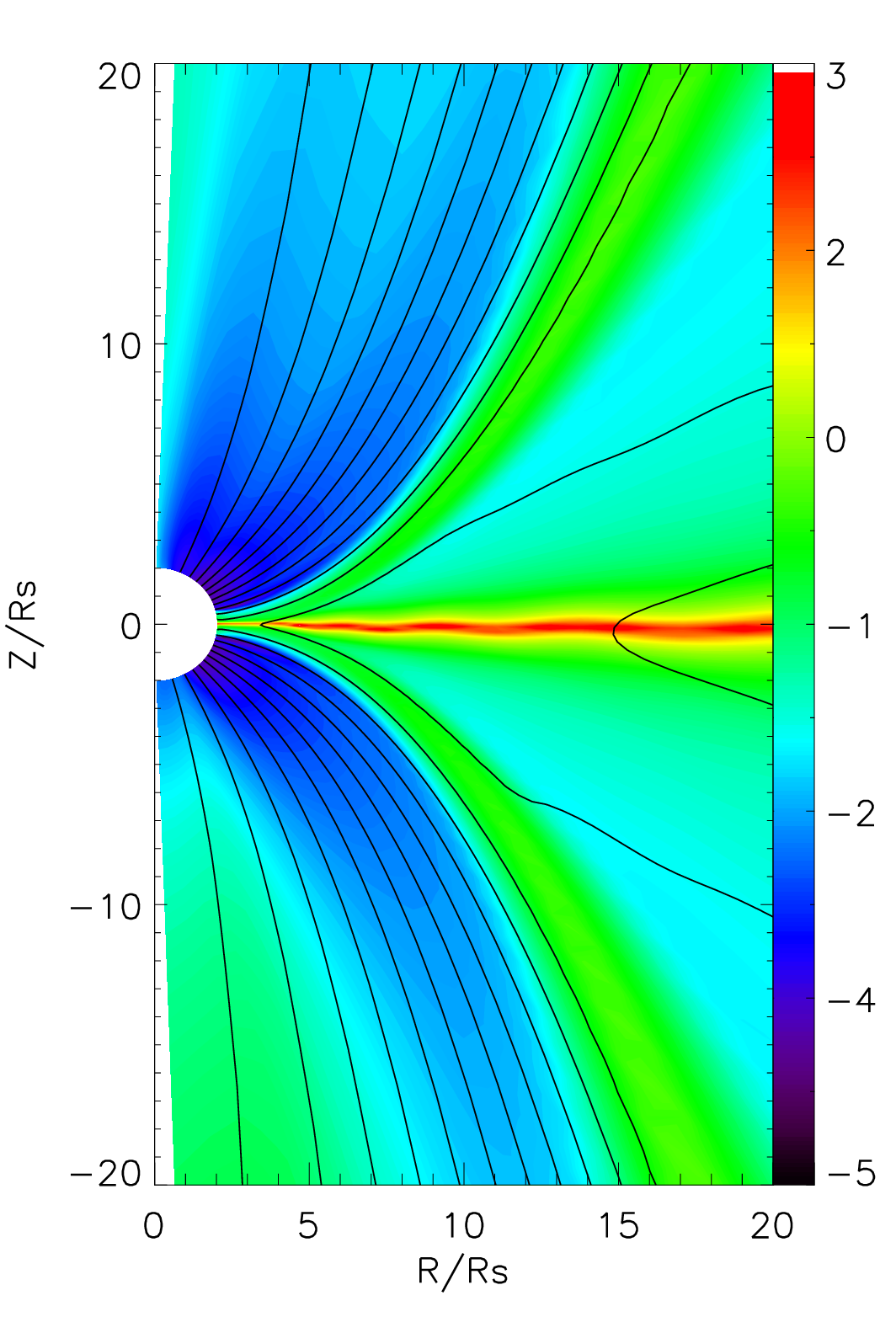}\hspace*{0.5cm}
\includegraphics[scale=0.28]{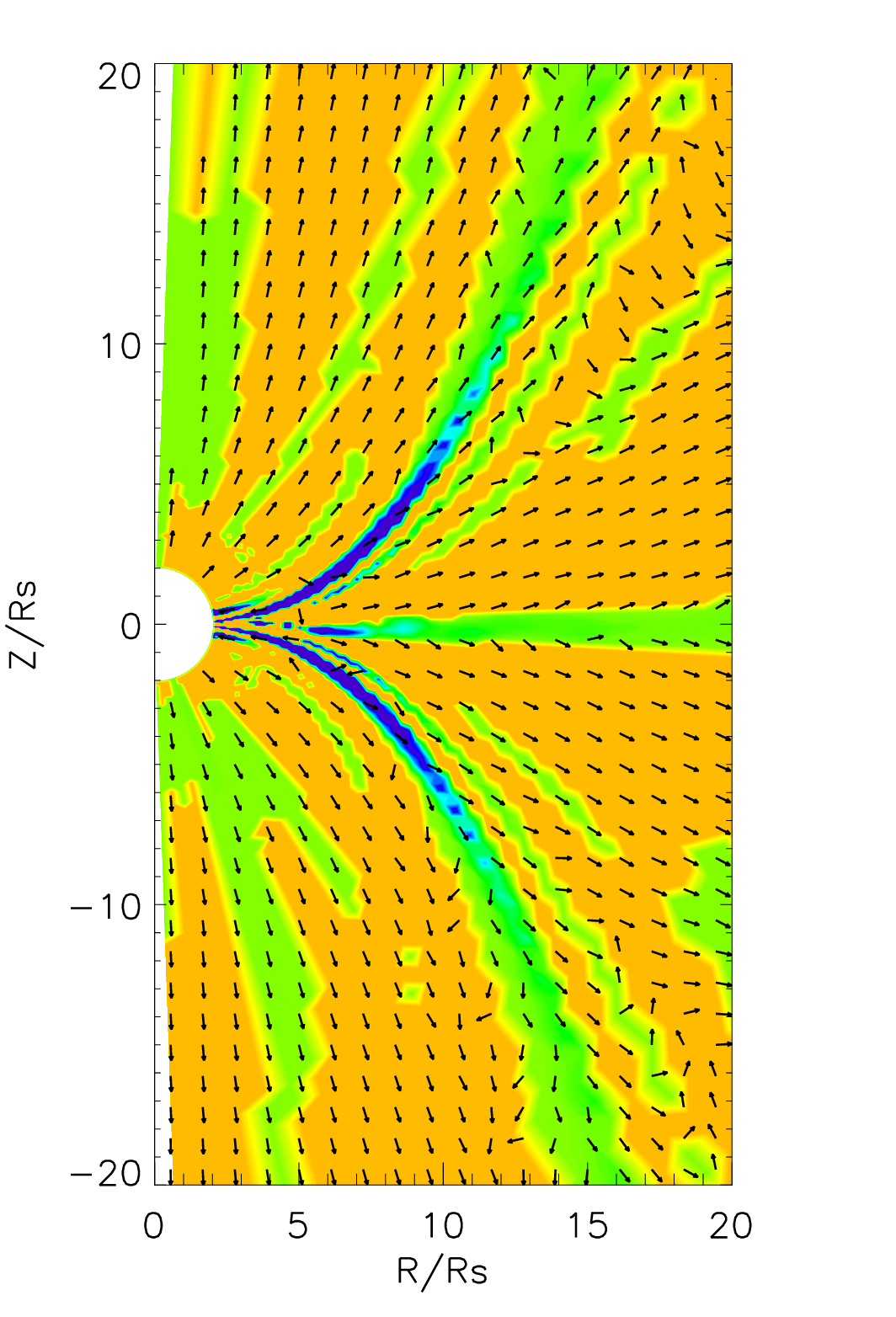}\\
\includegraphics[scale=0.28]{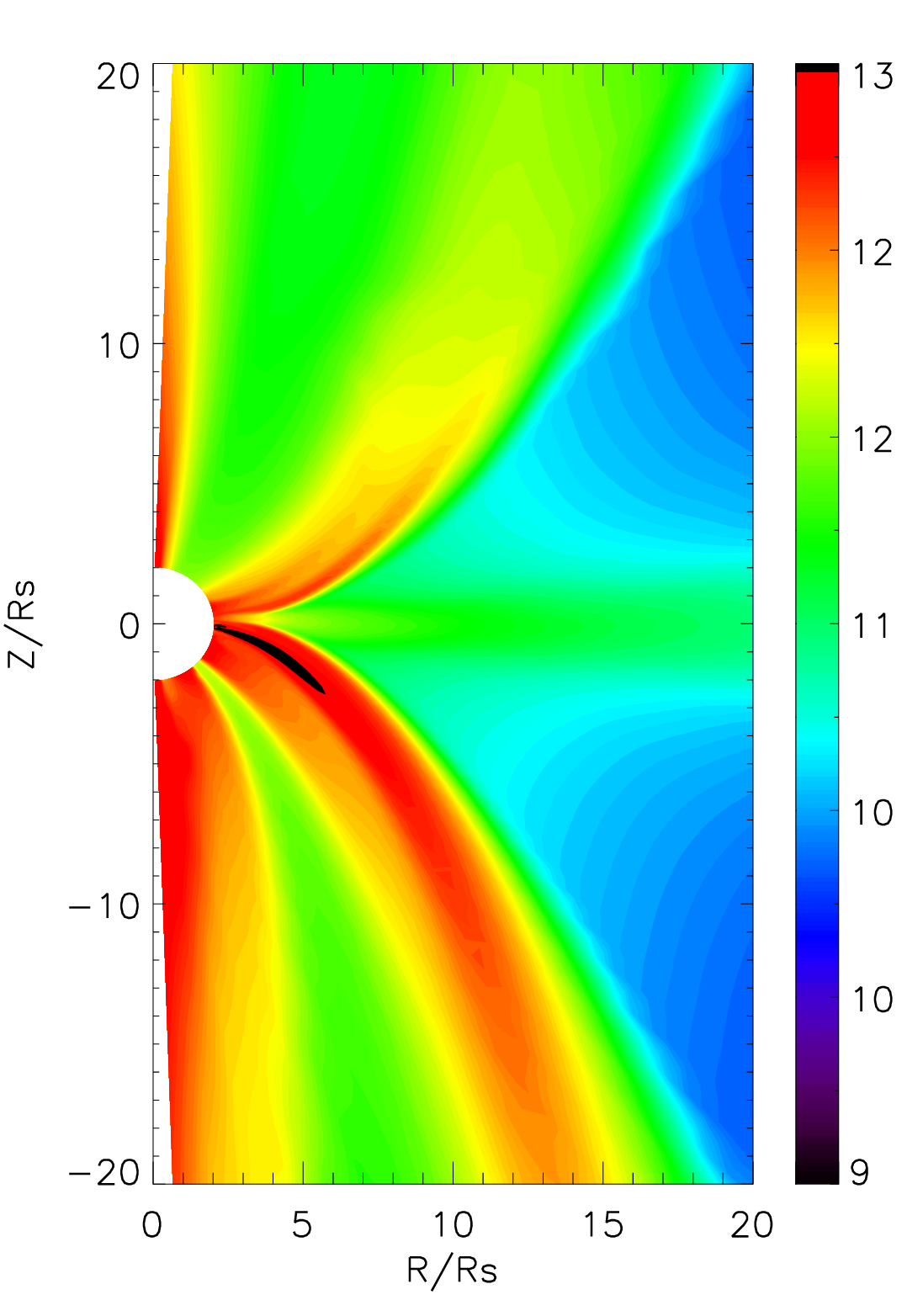}\hspace*{0.5cm}
\includegraphics[scale=0.28]{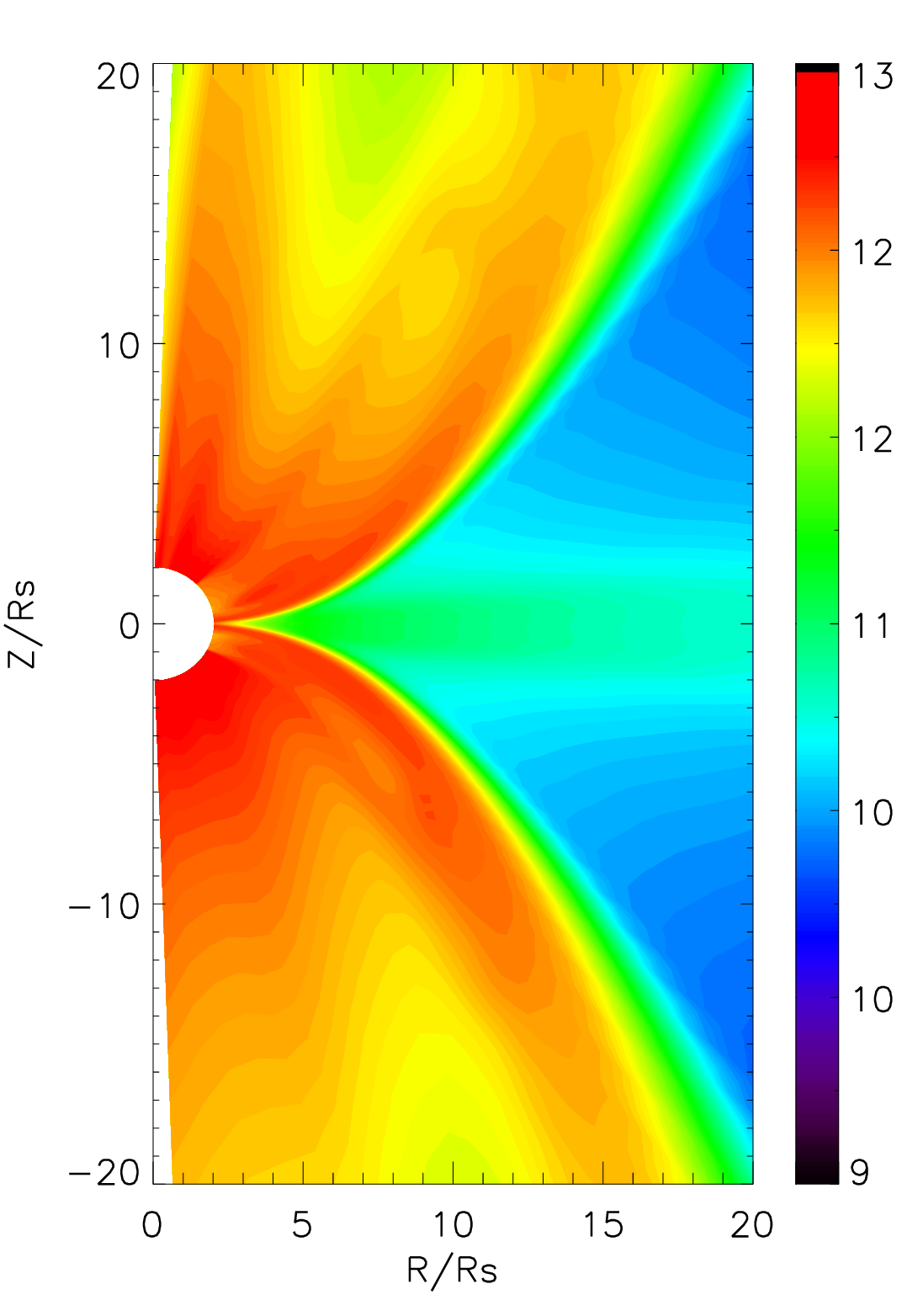}

\caption{Upper-left panel shows the snapshot of logarithm plasma
$\beta=p_{\rm gas}/p_{\rm mag}$ (colors) and magnetic field lines
(solid line) of model A1. Upper-right panel shows the snapshot of
unit vector of conduction energy flux $\bf Q / |\bf Q|$ (arrows) and
divergence of conduction flux $\nabla \cdot \bf Q$ (colors) of model
A1, note that orange color denotes region where gas is heated by
thermal conduction and green color denotes region where gas is
cooled by thermal conduction. Bottom panels show the time-averaged
logarithm temperature for models A0 (left) and A1 (right).}
\label{Fig:conductionA}
\end{center}
\end{figure}

In order to figure out why the inflowing gas becomes wind, we plot
magnetic field, heat flux and temperature in Figure
\ref{Fig:conductionA}. In Figure \ref{Fig:conductionA}, upper-left
panel shows the snapshot of logarithm plasma $\beta$ and magnetic
field lines for model A1. Black lines are magnetic field lines;
colors show logarithm plasma $\beta$. Upper-right panel shows the
snapshot of conduction energy flux $\bf Q$ and its divergence for
model A1; colors show the divergence of conduction flux $\nabla
\cdot \bf Q$ for model A1. Arrows show the energy flux. Orange
denotes regions those are heated by thermal conduction; green
denotes regions those are cooled by thermal conduction. Bottom
panels show the time-averaged (from $t=1.5$ to 1.6 orbits) logarithm
temperature for models A0 (left) and A1 (right), respectively. In
model A series, because of a relative small plasma $\beta$ and
dipolar configuration field as the initial condition, the wavelength
of the magneto-rotational instability (MRI; Balbus \& Hawley 1991;
1998) is large which suppressed the turbulence driven by the MRI.
Thus, the magnetic field evolves to a relative strong and ordered
magnetic field as the upper-left panel of Figure
\ref{Fig:conductionA} shows. The thermal conduction flux is mainly
along magnetic field lines, so it is also become very ordered. Heat
energy is transferred from inner to outer region by thermal
conduction. At the equator, the heat is transport from inner to
larger radius and plays a cooing role. It is clear that gas in the
regions $26^\circ <\theta < 45^\circ $ and $135^\circ <\theta <
154^\circ$ is heated up by the conduction flux along magnetic field
lines. The increase of temperature in those regions results in the
increase of gas pressure gradient force. Consequently, the increased
gas pressure gradient force makes the inflowing gas changing moving
direction and becoming wind. From the energy point of view, the
Bernoulli parameter of the gas in these two regions becomes larger
due to thermal conduction thus the gas is easier to become wind.
This argument of wind production is similar to that presented in Gu
(2015).

\begin{figure}
\begin{center}
\includegraphics[scale=0.45]{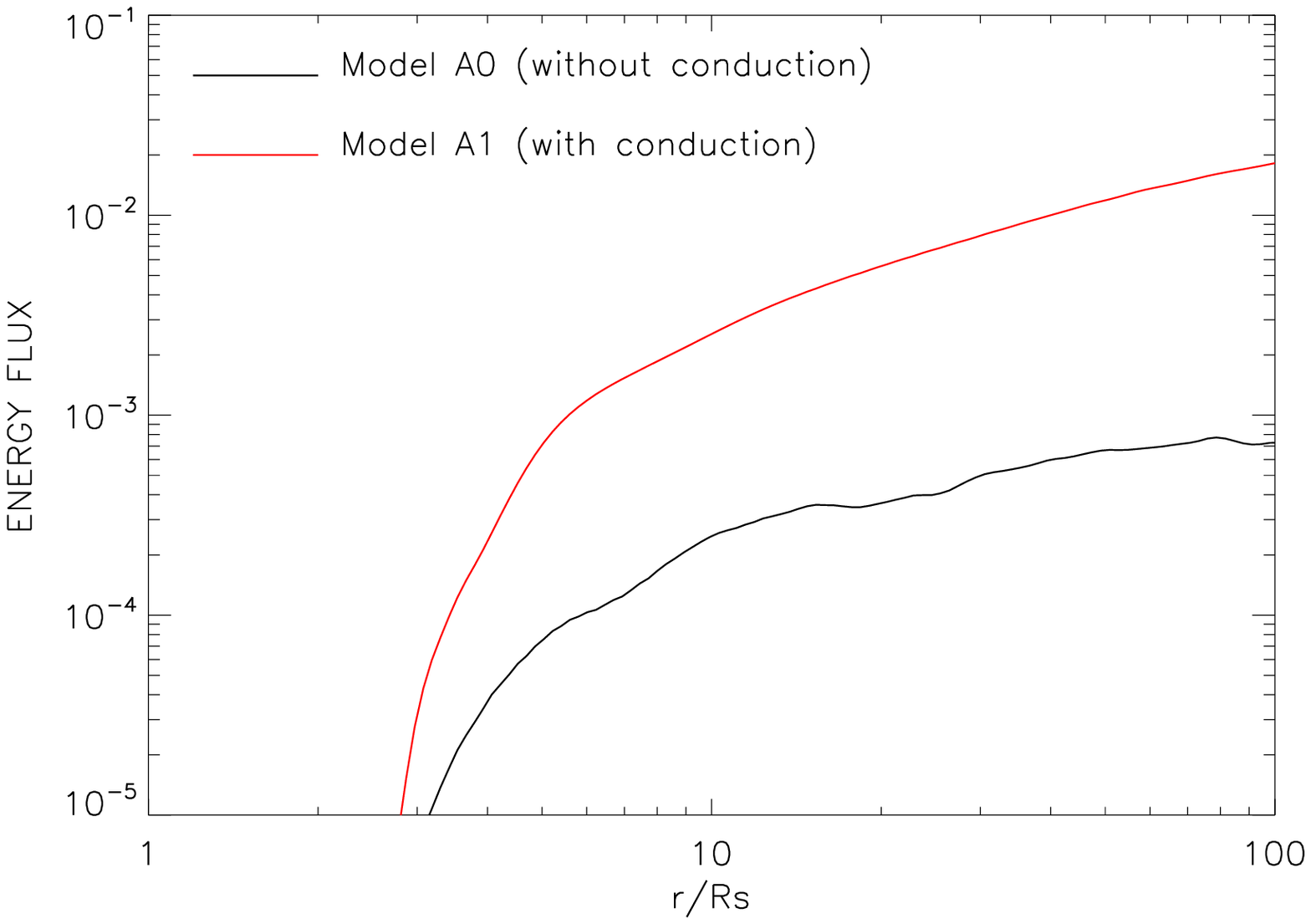}
\includegraphics[scale=0.45]{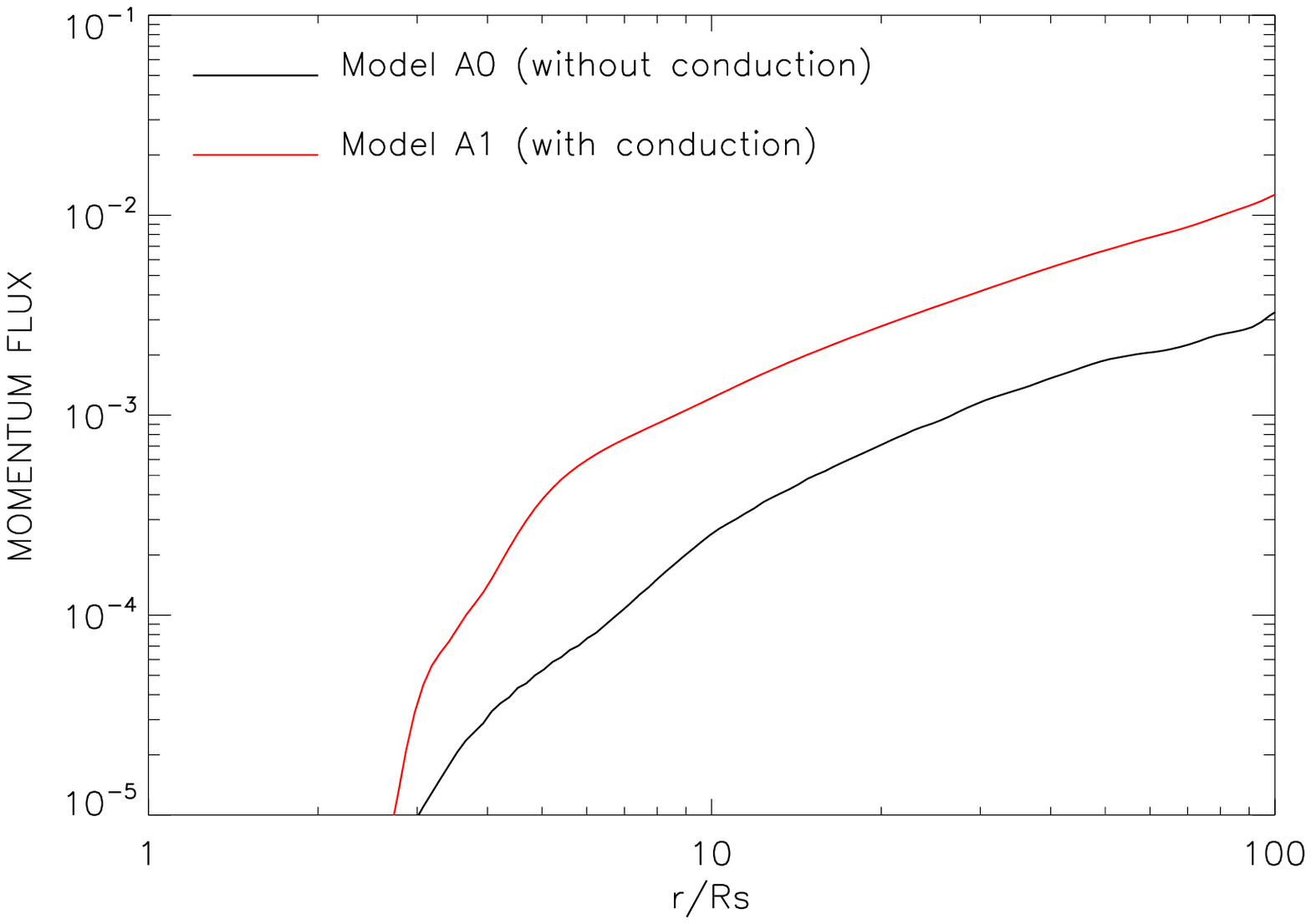}
\caption{Upper panel: Radial profile for energy fluxes (see Equation
(\ref{windenergy})) carried by wind for model A0 (black line) and A1
(red line). Lower panel: Radial profile for momentum fluxes (see
Equation (\ref{windmomentum})) carried by wind for model A0 (black
line) and A1 (red line). }\label{Fig:energyfluxA}
\end{center}
\end{figure}

\begin{figure}
\begin{center}
\includegraphics[scale=0.50]{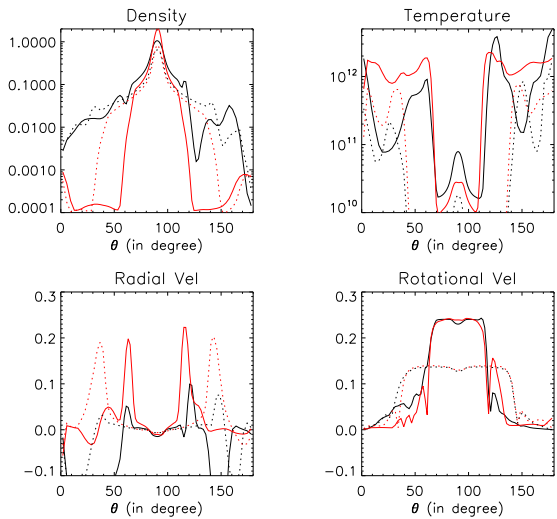}

\caption{Angular profiles of a variety of time-averaged variables of
model A0 (black lines, without conduction) and A1 (red lines, with
thermal conduction) at $r=10\rm R_s$ (solid lines) and 50 $\rm R_s$
(dotted lines).  From left to right, upper to bottom, the panels
denote the density, temperature, radial velocity and angular
velocity, respectively.}\label{Fig:thetaA}
\end{center}
\end{figure}

\begin{figure}
\begin{center}
\includegraphics[scale=0.5]{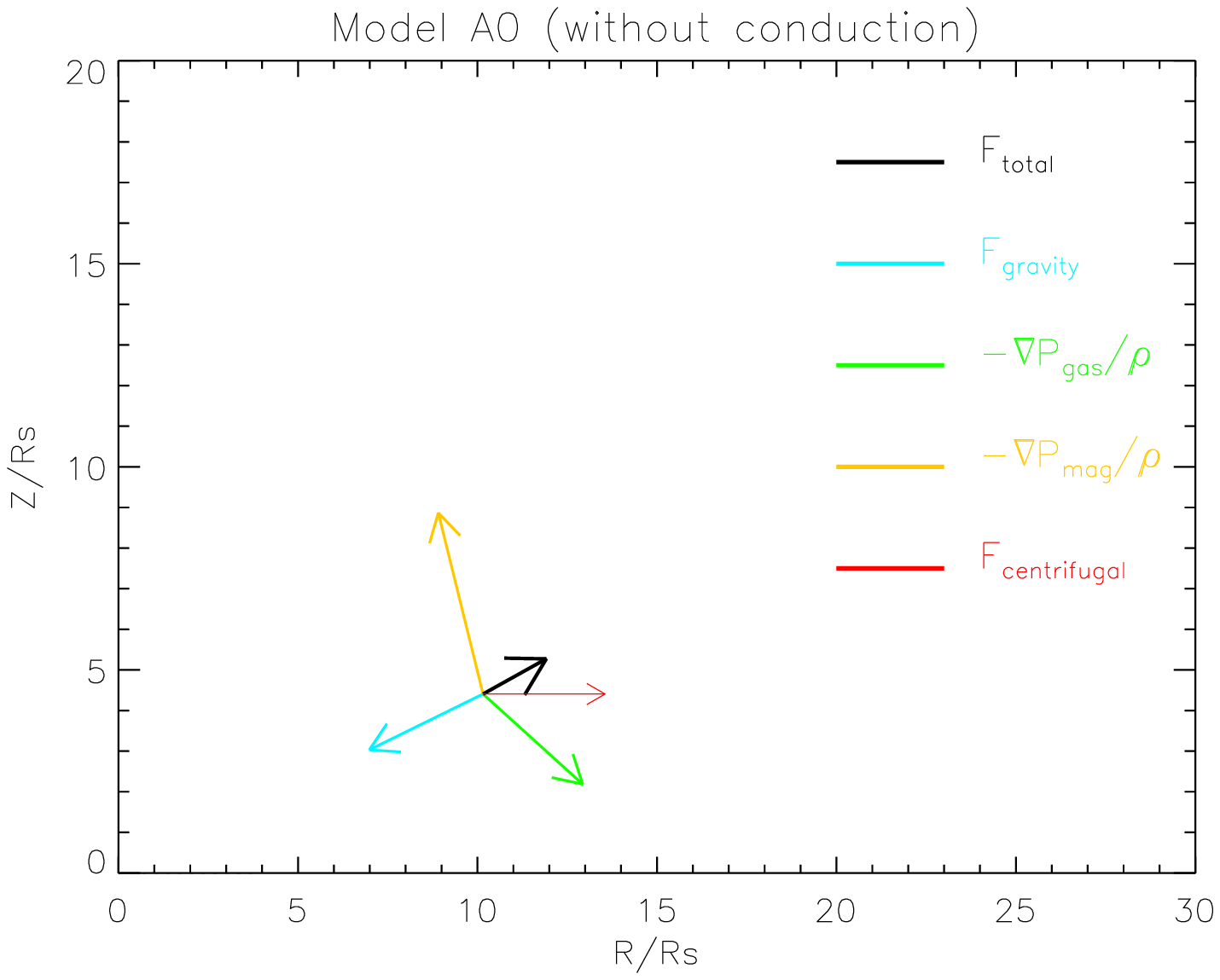}
\includegraphics[scale=0.5]{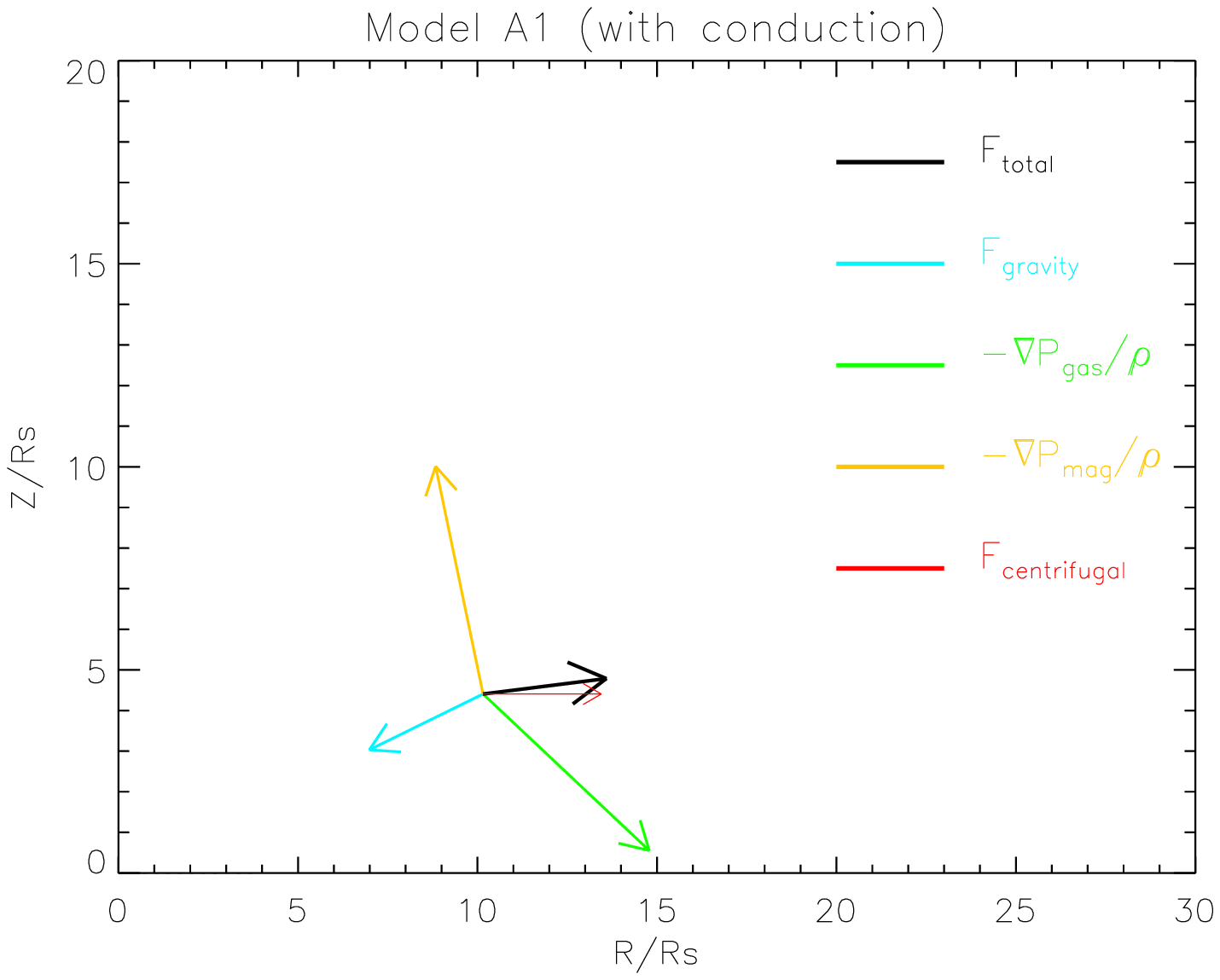}

\caption{Force analysis at $r=10R_s$ and $\theta=65^\circ$ for model
A0 (upper-panel) and A1 (lower-panel). The arrows indicate force
direction, whose length represents force magnitude. All the forces
are scaled in the same way, the scaled factor chosen arbitrarily to
fit the arrows in both models. }\label{Fig:forceA}
\end{center}
\end{figure}

Furthermore, to investigate how the nature of outflow changed by
thermal conduction, we also calculate the outflow energy fluxes
carried out in the form of the kinetic energy and outflow momentum
fluxes as follows:
\begin{equation}
\dot{E}_{wind}(r)= 2\pi r^2 \int_{0}^{\pi} \frac{1}{2} \rho \max
(v_r, 0) v_r^2 \sin \theta \rm d\theta, \label{windenergy}
\end{equation}
\begin{equation}
\dot{P}_{wind}(r)= 2\pi r^2 \int_{0}^{\pi} \rho \max (v_r, 0) v_r
\sin \theta \rm d\theta. \label{windmomentum}
\end{equation}
Figure \ref{Fig:energyfluxA} shows the energy and momentum fluxes
carried by wind. From this figure, it is clear that in both models
the energy and momentum fluxes carried by winds increase outward.
This is consistent with Yuan et al. (2015). Compared with model A0,
the energy flux in model A1 increases by about one order of
magnitude. In order to study why the energy and momentum fluxes
increase when thermal conduction is included, we plot Figure
\ref{Fig:thetaA}, which shows the angular distributions of various
quantities. Let us take the physical values for wind at $10R_{\rm
S}$ as an example. At $10R_{\rm S}$, the wind mainly occurs around
$\theta=60^\circ-70^\circ$. After considering thermal conduction, in
the wind region, the density  decreases by a factor $\sim 5$; but
the velocity increases by a factor of $\sim 4$. Since
$\dot{E}_{wind} \propto \rho v_r^3$ and $\dot{P}_{wind} \propto \rho
v_r^2$, the energy flux of wind increases by a factor of $\sim 10$,
the momentum flux by a factor of $\sim 3$.

To figure out the reason of why the velocity of wind increases after
the inclusion of thermal conduction, we calculate the forces
exerting on gas. Based on the GRMHD simulation data, Yuan et al.
(2015) (see also Moller \& Sadowski 2015) have shown that the wind
is mainly driven by the centrifugal force, the gradient force of gas
pressure and magnetic pressure. Figure \ref{Fig:forceA} plots the
driving force for wind at $\theta=65^\circ$ and $r=10R_{\rm S}$ in
model A0 (upper panel) and A1 (lower panel). In the case of model
A0, the main driving forces are the combination of the gradient
force of magnetic pressure and the gas pressure and the centrifugal
force. It is clear that after the thermal conduction is included the
gas pressure gradient force increases due to the increase of
temperature. The magnetic pressure gradient force also increases.
The increase of forces results in the larger velocity of wind in
model A1 compared with A0.

\subsection {Quadrupolar field models}
In model A series, the magnetic field is strong and ordered.
Therefore, thermal conduction along magnetic field can transport
energy from the inner to outer region and affects the wind
production significantly. If the magnetic field is tangled one may
expect that energy can not be transferred to large distance. In this
case, the effects of conduction may be  smaller compared with the
case when the magnetic field is ordered. In order to examine this
point, we carry out model B series (see Table 1). In this series,
the initial magnetic field configuration is quadrupolar (see the
right panel of Figure \ref{Fig:field}). In this case, magnetic
reconnection very easily occurs, which can decrease the strength of
magnetic field. In this section, the data are usually time averaged
from $t= 3.0$ to $t= 3.5$ orbits, except for the snapshot data is at
time $t=3.5$ orbits.

\begin{figure}
\begin{center}
\includegraphics[scale=0.28]{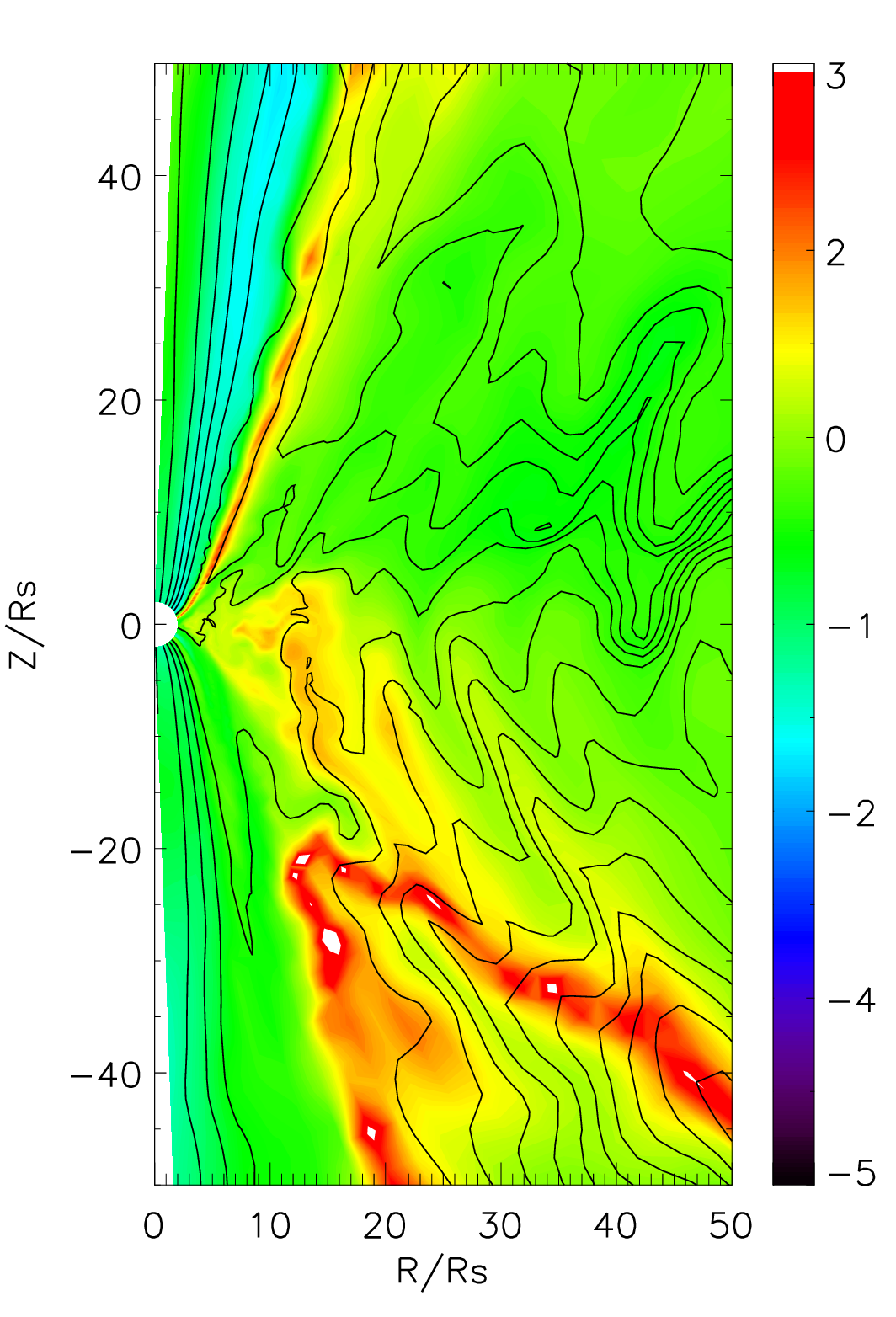}\hspace*{0.5cm}
\includegraphics[scale=0.28]{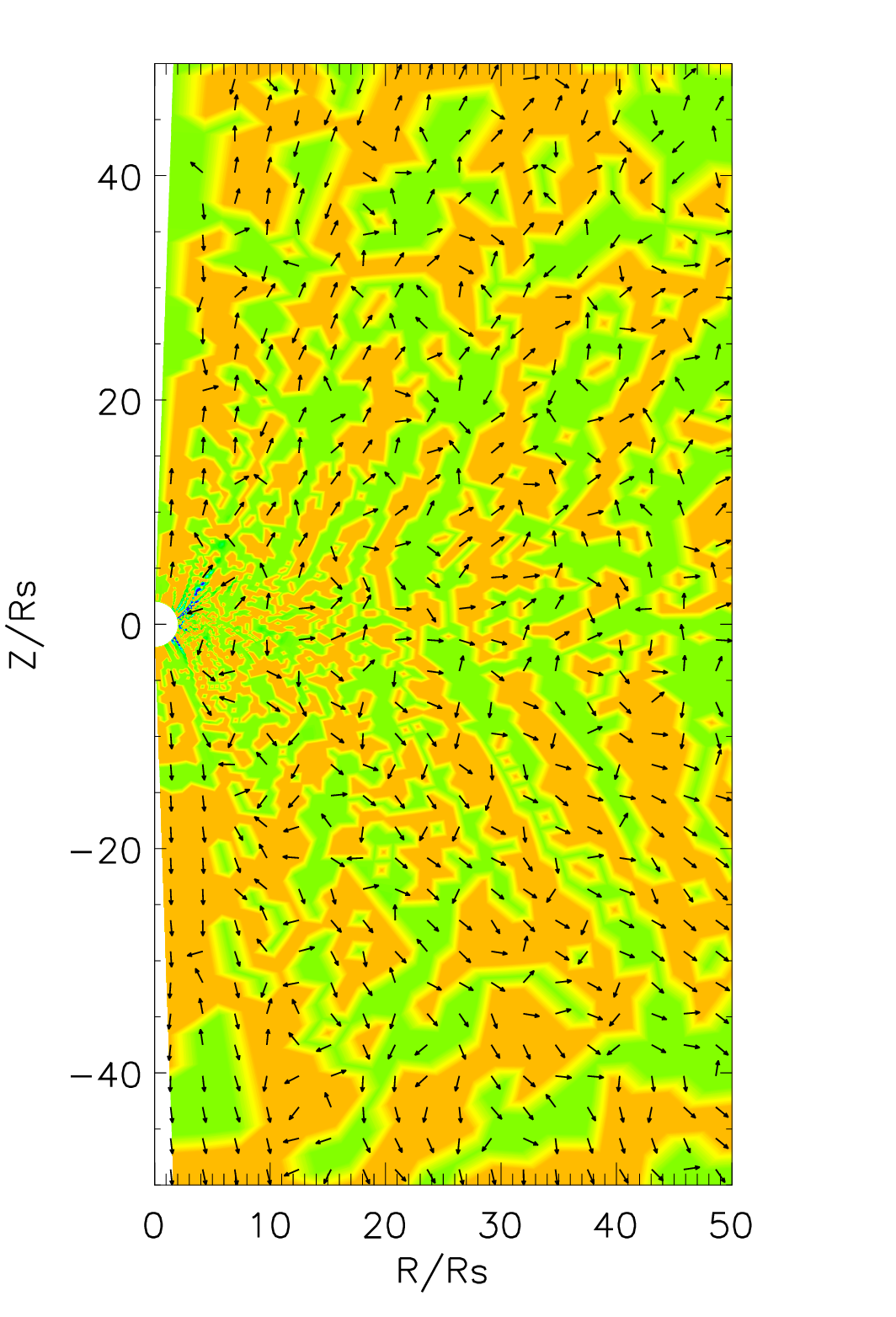}

\caption{Left panel shows the snapshot of logarithm plasma $\beta$
and magnetic field lines for model B1. Black lines are magnetic
field lines; colors show logarithm plasma $\beta$. The right panel
show the unit vector of conduction energy flux $\bf Q / |\bf Q|$
(arrows) and the divergence of conduction flux $\nabla \cdot \bf Q$
(colors) for model B1, Note that in right panel, orange color
denotes region which gas is heated by thermal conduction; green
color denotes region which gas is cooled by thermal
conduction.}\label{Fig:plasmaq}
\end{center}
\end{figure}

Figure \ref{Fig:plasmaq} shows the snapshot of logarithm of $\beta$
over-plotted with magnetic field lines (left panel) and the unit vector of
conduction energy flux $\bf Q / |\bf Q|$ over-plotted with the divergence of
conduction flux $\nabla \cdot \bf Q$ (right panel) of model B1. Recall that
in model A series, the magnetic field is ordered and the heat flux is transported mainly along the magnetic field
lines so the heat flux is also ordered as seen in the upper-right panel
of Figure \ref{Fig:conductionA}. But the case is different in model
B series. Because  the magnetic field is weaker and
tangled, as shown in Figure \ref{Fig:plasmaq} (left panel, solid
line), the conduction energy flux (right panel,
arrows) is thus not ordered.

\begin{figure}
\begin{center}
\includegraphics[scale=0.45]{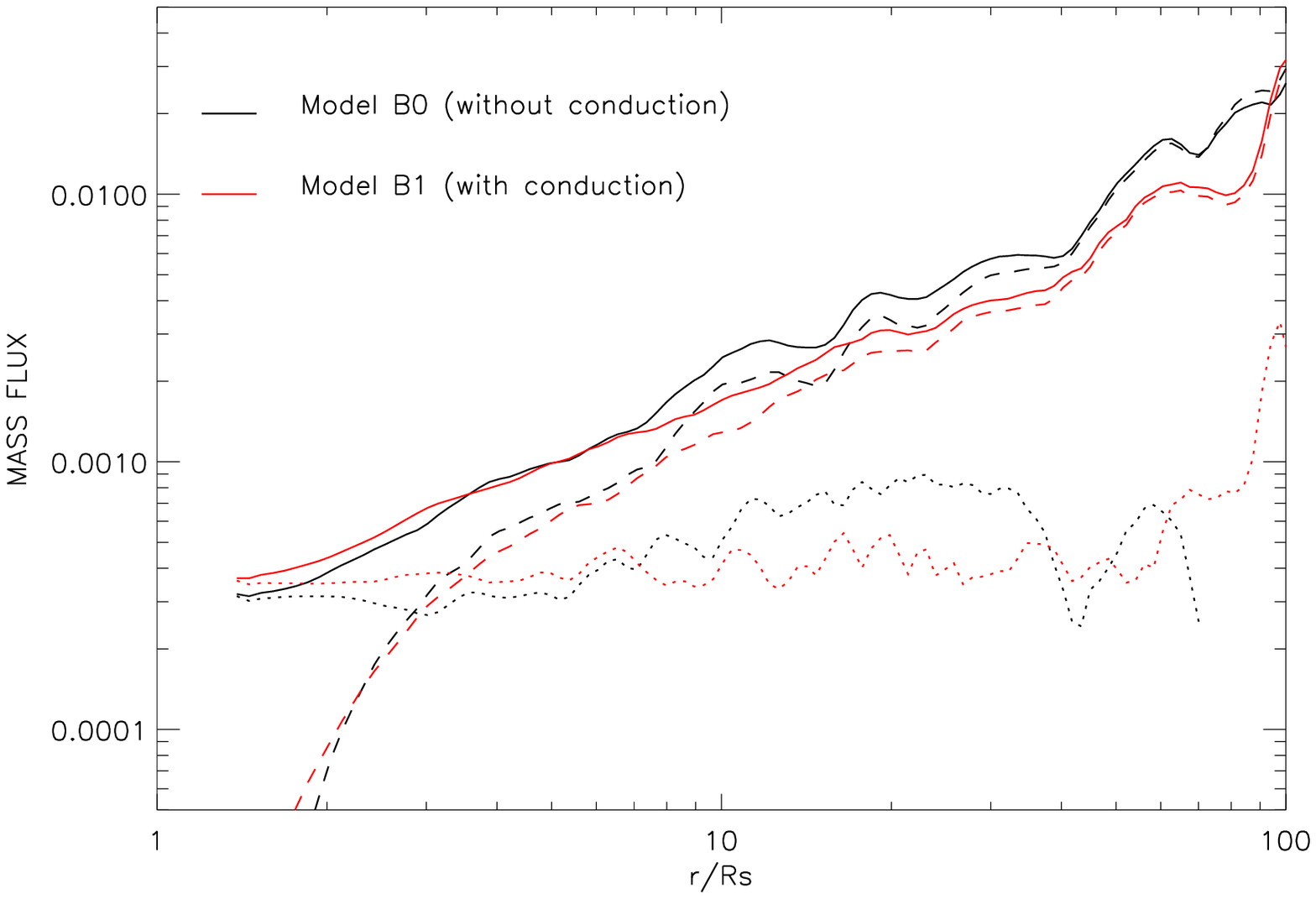}
\caption{ The radial profiles of the time-averaged and angle
integrated mass inflow rate $\dot{M}_{\rm in}$ (solid line), outflow
rate $\dot{M}_{\rm out}$ (dashed line), and the net rate
$\dot{M}_{\rm acc}$ (dotted line) in model B0 (black lines) and B1
(red lines). \label{Fig:accretionrateB}}
\end{center}
\end{figure}

\begin{figure}
\begin{center}
\includegraphics[scale=0.5]{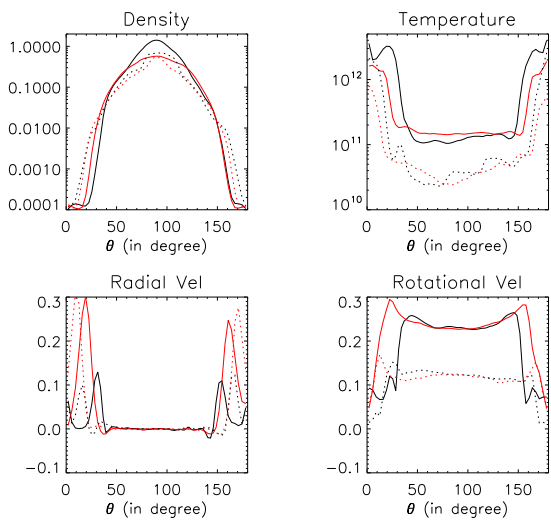}
\caption{Angular profiles of a variety of time-averaged variables
for model B0 (black lines, without conduction) and B1 (red lines,
with thermal conduction) at $r=10\rm R_s$ (solid lines) and 50 $\rm
R_s$ (dotted lines).}\label{Fig:theta_q}
\end{center}
\end{figure}
\begin{figure}
\begin{center}
\includegraphics[scale=0.45]{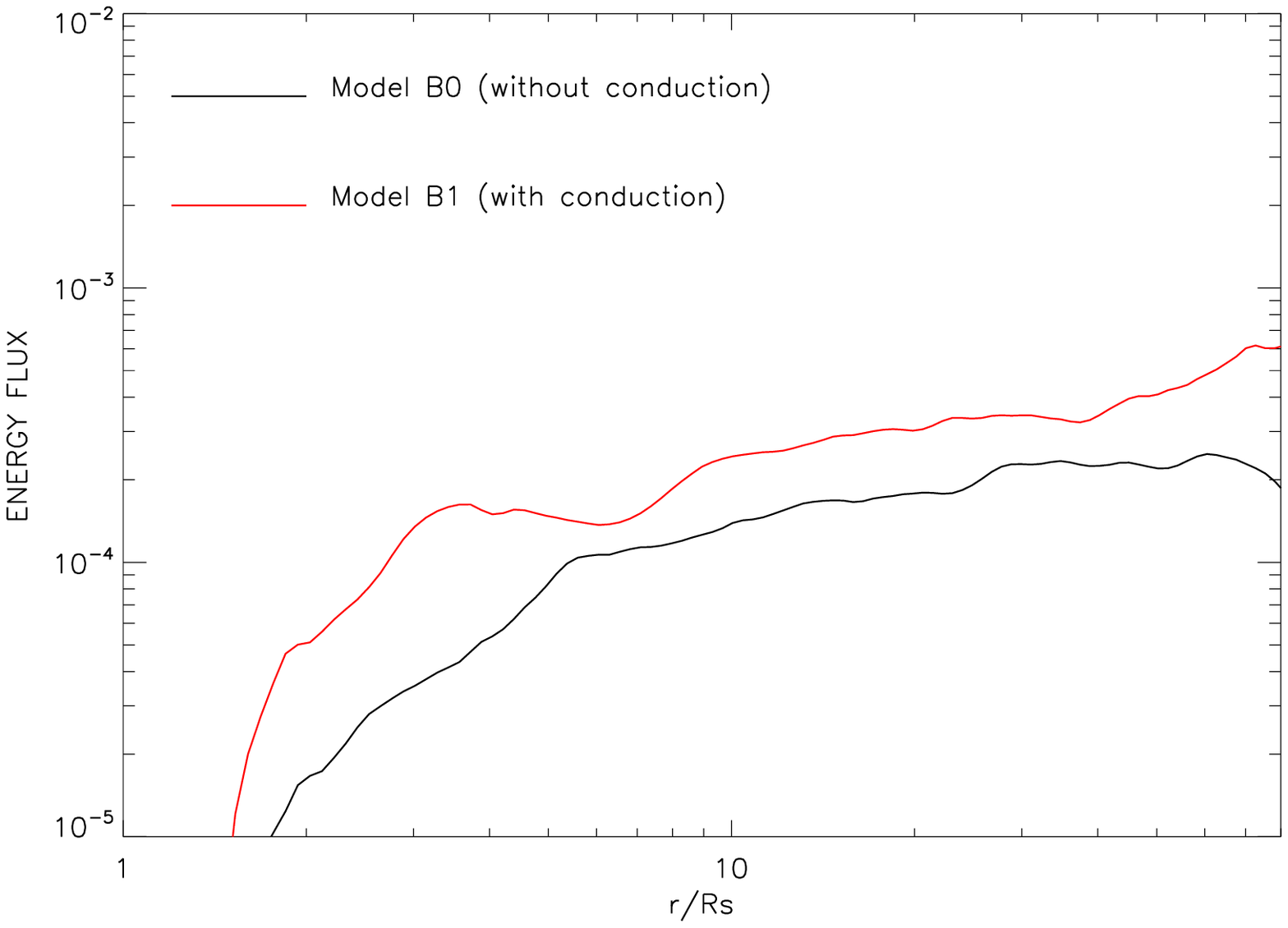}
\includegraphics[scale=0.45]{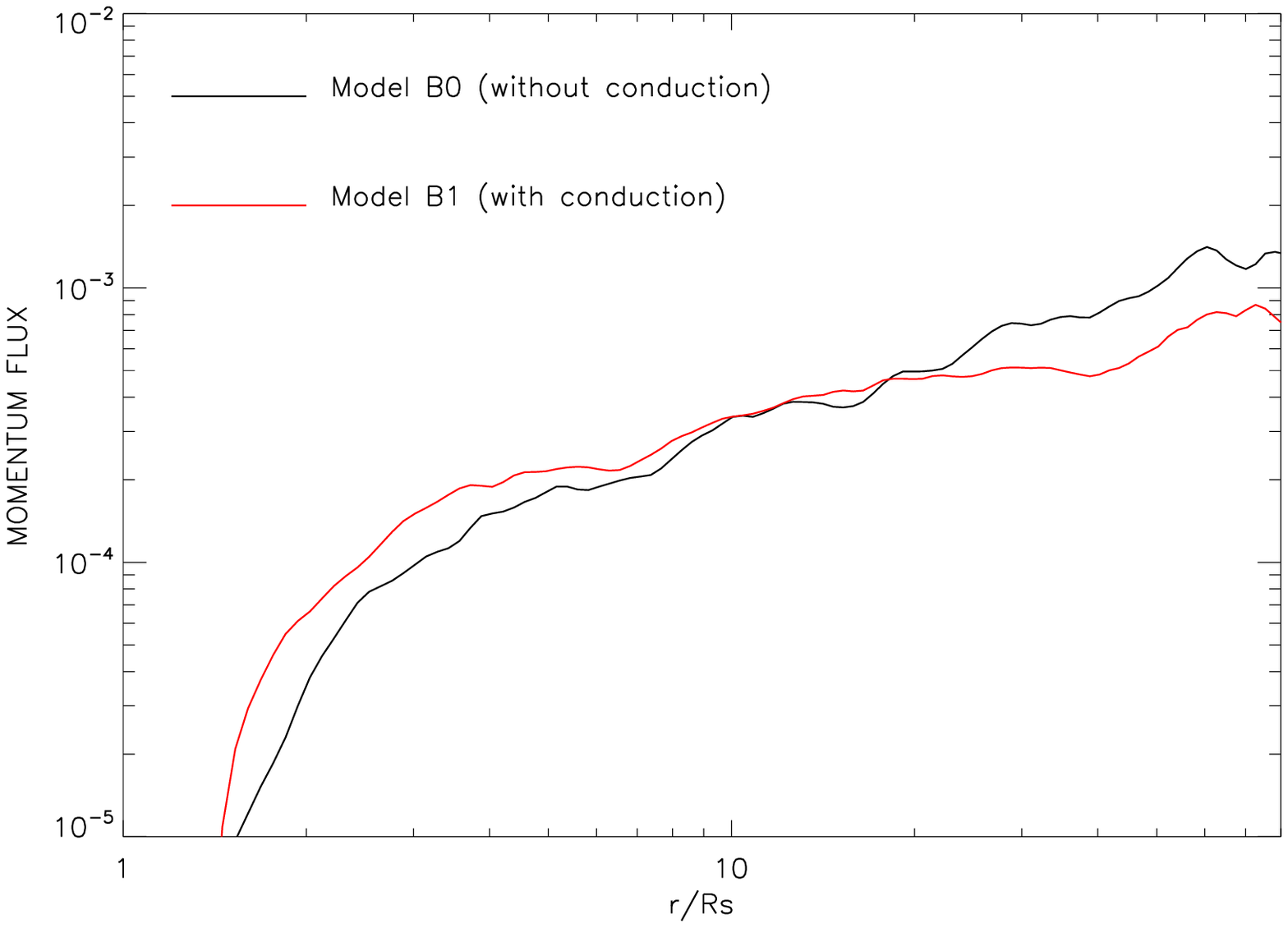}
\caption{Upper panel: Radial profile for energy fluxes (see Equation
(\ref{windenergy})) carried by wind for model B0 (black line) andB1
(red line). Lower panel: Radial profile for momentum fluxes (see
Equation (\ref{windmomentum})) carried by wind for model B0 (black
line) and B1 (red line). From left to right, upper to bottom, the
panels denote the density, temperature, radial velocity and angular
velocity, respectively. }\label{Fig:energyfluxB}
\end{center}
\end{figure}
\begin{figure}
\begin{center}
\includegraphics[scale=0.5]{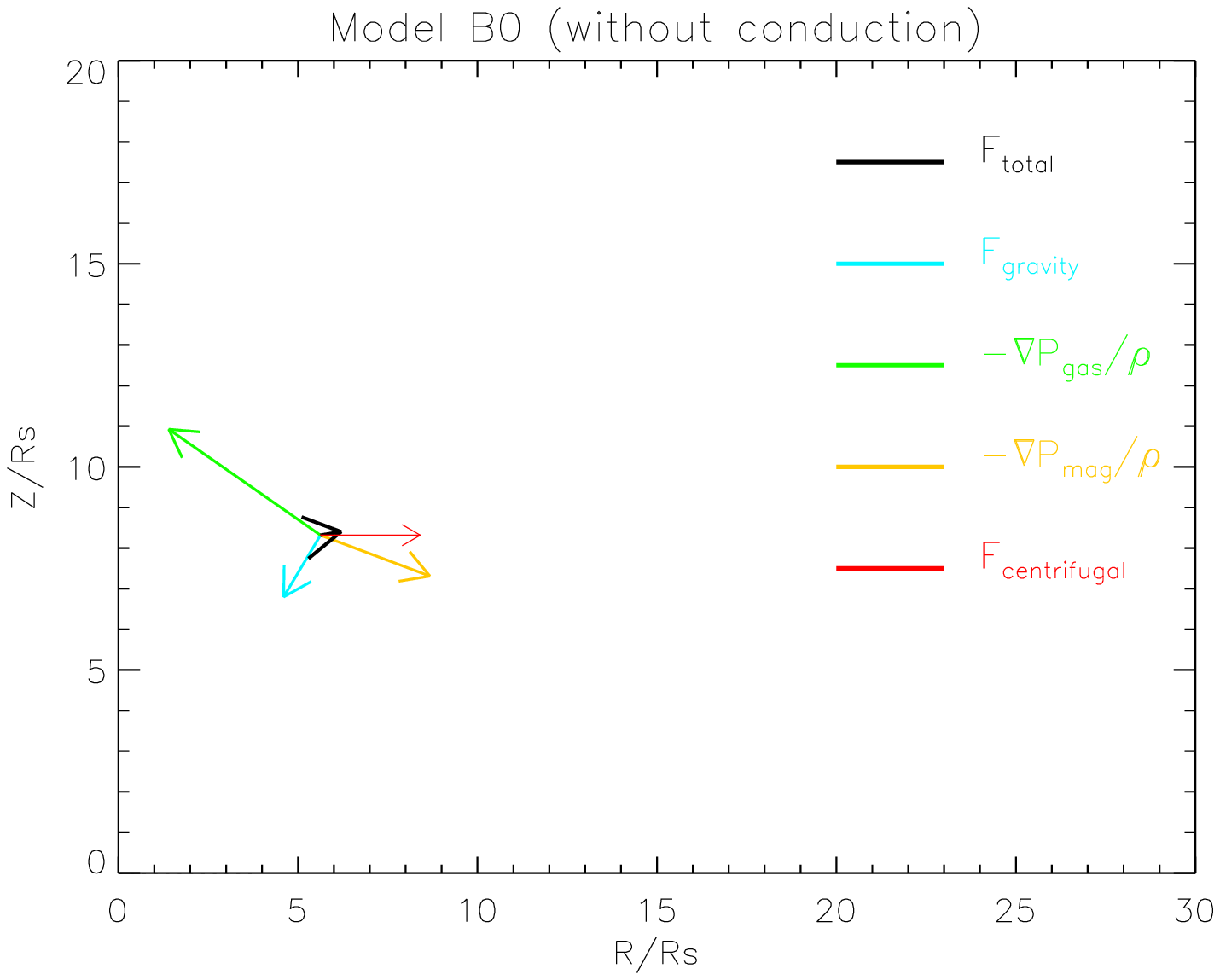}
\includegraphics[scale=0.5]{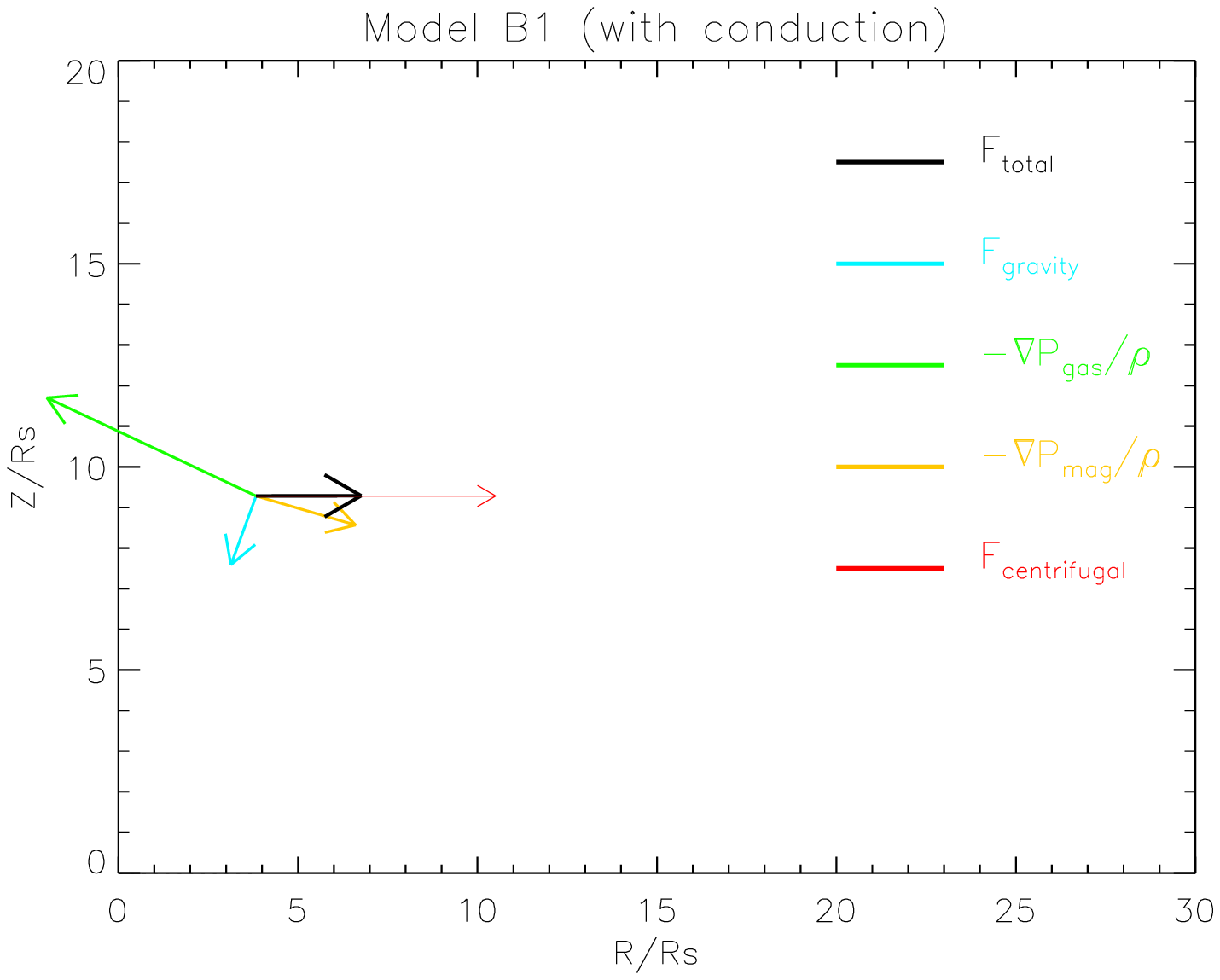}
\caption{Force analysis at radius of $10R_{\rm s} $ for model B0
(upper-panel, $\theta=30^\circ$) and  model B1 (lower-panel,
$\theta=20^\circ$) }\label{Fig:forceB}
\end{center}
\end{figure}
Figure \ref{Fig:accretionrateB} shows the mass accretion rates in
model B series. It is clear that the mass accretion rates are only
slightly changed after thermal conduction is taken into account. In
order to study the reason, we plot the angular profiles of some
quantities in Figure \ref{Fig:theta_q}. As shown in the upper-right
panel, the angular profiles of temperature in model B1 is flatter
than that in model B0, consistent with that found in Johnson \&
Quataert (2007). Except for the changes of density and temperature,
the radial and rotational velocities of wind are also changed when
considering thermal conduction. Now, we make a quantitative analysis
of the changes of wind mass fluxes based on the changes of density
and velocity of wind in model B1. Taking the physical values for
wind at $10R_s$ as an example, wind is present at $\theta=30^\circ$
in model B0 and $\theta=20^\circ$ in model B1. When conduction is
included, the velocity of wind increases but the density of wind
decreases. So the mass flux of wind is just slightly decreased by a
factor of $\sim 2$.

Figure \ref{Fig:energyfluxB} shows the energy and momentum fluxes
carried by wind. It is clear that the momentum flux of wind is just
slightly changed when thermal conduction is included. But the energy
flux increases by a factor of $\sim 2$. This is because that the
momentum and energy fluxes of wind are $\dot{P}_{wind} \propto \rho
v_r^2$ and $\dot{E}_{wind} \propto \rho v_r^3$, respectively. In
model B series, when conduction is included, the density of wind is
decreased by a factor of $\sim 5$ and velocity of wind is increased
by a factor of $\sim 2$. Therefore, the momentum flux is almost
unchanged but the energy flux is increased by a factor of 2.

Following previous analysis of wind, in Figure \ref{Fig:forceB}, we
plot the driving forces for wind in model B series. In both models,
the main driving forces of wind are the combination of the gradient
of magnetic pressure and gas pressure and the centrifugal force. The
gas pressure gradient force is increased after conduction is
considered consistent with the results in model A series. The
centrifugal force in model B1 is two times of that in model B0. The
increase of centrifugal force is because that in model B1, the
rotational velocity of wind is larger than that in model B0 (see the
lower-right panel of Figure \ref{Fig:theta_q}).

\section{Summary and discussion}
Previous works have shown that strong winds exist in hot accretion
flows (Yuan et al. 2015; see also Yuan et al. 2012b; Narayan et al.
2012; Li et al. 2013)\footnote{Narayan et al. (2012) find that the
wind is very weak. The reason for the discrepancy has been analyzed
in Yuan et al. (2015).}. Those works do not include thermal
conduction. In extremely low accretion rate systems such as the
accretion flow in Galactic center Sgr $\rm A^*$, the plasma is very
dilute, and the collisional mean-free path of electrons is large and
much greater than their gyro-radius. Thus, thermal conduction is
dynamically important and anisotropic, along magnetic field lines.
In this paper, we have studied the effects of anisotropic thermal
conduction on the wind by performing two-dimensional MHD
simulations. Two different magnetic field topologies are considered:
a strong ordered field and a weaker tangled field.  Our simulation
results show that thermal conduction has moderate effects on the
mass flux of wind in both cases. However, the energy flux of wind
can be increased by a factor of $10$ due to the increase in wind
velocity when thermal conduction is included in ordered magnetic
field case. The increase of wind velocity is because the driving
forces (e.g. gas pressure gradient force and centrifugal force)
increase when thermal conduction is included.

There are still some limitations in our work. Our simulations are
two-dimensional and we adopt the one-temperature simplification.
Ressler et al. (2015) study the two-temperature hot accretion flow
with anisotropic conduction in GRMHD. Their results have shown that
electron heating rates depend on the local magnetic field strength
and electron thermal conduction modifies the electron temperature in
the inner regions of accretion flows. They use a quadrupolar
magnetic field and our model B1 is basically consistent with their
results, such as the temperature gradient is flatter due to the
effects of thermal conduction. But they did not give the results of
wind properties.

Finally, we would like to mention that in a weakly collisional
accretion flow, the ion mean-free path can be much greater than its
gyro-radius, and thus the pressure tensor is anisotropic. In this
case, the growth rate of the MRI can increase dramatically at small
wave numbers compared with  MRI in ideal MHD (Quataert et al. 2002;
Sharma et al. 2003). In this regime, the viscous stress tensor is
anisotropic, and Balbus (2004) has shown that when anisotropic
viscosity is included, the flow is subject to the magnetoviscous
instability (MVI; see also Islam \& Balbus 2005). An interesting
project in the future is to investigate the effects of MVI on wind
from hot accretion flows.

\section*{Acknowledgments}
We thank the referee for rasing very useful questions, which help us
to improve this paper significantly. We thank Feng Yuan and Fu-Guo
Xie for the valuable discussions. De-Fu Bu is supported in part by
the National Basic Research Program of China (973 Program, grant
2014CB845800), the Strategic Priority Research Program ¡°The
Emergence of Cosmological Structures¡± of the Chinese Academy of
Sciences (grant XDB09000000), and the Natural Science Foundation of
China (grants 11103061, 11133005, 11121062, and 11573051). Mao-Chun
Wu is supported in part by the Natural Science Foundation of China
(grants U1431228, 11133005, 11233003, 11421303), the National Basic
Research Program of China (2012CB821801), the Strategic Priority
Research Program of the Chinese Academy of Sciences (XDB09000000),
and the grant from ``the Fundamental Research Funds for the Central
Universities''. This work made use of the High Performance Computing
Resource in the Core Facility for Advanced Research Computing at
Shanghai Astronomical Observatory.

\label{lastpage}

\end{document}